\documentclass[10pt,onecolumn,aps,prd,preprintnumbers,showpacs,superscriptaddress,nofootinbib,amsmath,amssymb,floats,floatfix,showkeys,notitlepage,longbibliography]{revtex4-2}

\usepackage{orcidlink}
\usepackage{comment}
\usepackage{lipsum}
\usepackage{graphicx}
\usepackage{subfigure}
\usepackage{palatino}
\usepackage{sans}
\usepackage{hyperref}
\hypersetup{colorlinks=true,linkcolor=blue,urlcolor=blue,citecolor=blue}
\usepackage[toc,page]{appendix}
\usepackage[normalem]{ulem}
\usepackage{adjustbox}
\usepackage{latexsym}
\usepackage{amsmath}
\usepackage{amssymb}
\usepackage{amsfonts}
\usepackage{dcolumn}
\usepackage{bm}
\usepackage{tikz}
\usetikzlibrary{decorations.pathmorphing}
\usepackage{bigints}
\usepackage{array,tabularx,multirow,booktabs}
\usepackage[tracking=true]{microtype}
\usepackage{soul} 
\SetTracking{}{500}
\SetTracking{encoding={*}, shape=sc}{40}
\UseRawInputEncoding 
\allowdisplaybreaks

\usepackage[utf8]{inputenc}
\usepackage{xcolor} 

\renewcommand{\Re}{\ensuremath{\mathrm{Re}}}

\begin{document} \sloppy

\title{Shadow Ringing of Black Holes from Photon Sphere Quasinormal Modes}

\author{Reggie C. Pantig \orcidlink{0000-0002-3101-8591}} 
\email{rcpantig@mapua.edu.ph}
\affiliation{Physics Department, School of Foundational Studies and Education, Map\'ua University, 658 Muralla St., Intramuros, Manila 1002, Philippines.}

\begin{abstract}
The recent convergence of gravitational-wave (GW) observations and black hole imaging provides complementary probes of strong-gravity dynamics. While the black hole shadow is typically modeled as a static feature, a dynamically perturbed spacetime in its ringdown phase must induce temporal modulations in the shadow's apparent size and shape. We develop a theoretical framework within linear perturbation theory to investigate this shadow ringing effect for a Schwarzschild black hole. By modeling the geometry as a small, mode-selected quasinormal mode (QNM) perturbation, we treat the shadow boundary as an instantaneous separatrix of null geodesics. We derive a first-order, gauge-invariant mapping between the metric perturbation $h_{\mu\nu}$ and the displacement of the shadow boundary, $\delta R(\varphi,t)$. By perturbing the effective potential for null geodesics near the unstable photon sphere ($r=3M$), we derive mode-resolved transfer coefficients that quantify how the QNM imprints itself onto the shadow. We predict that the shadow boundary oscillates coherently at the QNM's real frequency $\omega_{\rm Re}$ with an exponential damping rate set by $|\omega_{\rm Im}|$. Furthermore, the azimuthal structure of the modulation encodes the spherical harmonic content $(\ell,m)$ of the driving QNM, providing a novel, geometric signature for QNM spectroscopy.
\end{abstract}


\maketitle

\section{Introduction} \label{sec1}
Black holes admit characteristic spacetime oscillations called quasinormal modes (or QNMs) that dominate the late-time ringdown following a dynamical perturbation. In gravitational-wave (GW) observations, these damped sinusoids encode the mass, spin, and, more broadly, the near-horizon geometry \cite{LIGOScientific:2016aoc,LIGOScientific:2016lio, Konoplya:2022pbc,Konoplya:2023hqb}. In parallel, very-long-baseline interferometry (VLBI) at millimeter wavelengths has inaugurated black hole imaging, with the Event Horizon Telescope (EHT) delivering horizon-scale structure and a silhouette commonly termed the shadow \cite{EventHorizonTelescope:2019dse,EventHorizonTelescope:2022wkp}, which has been long theorized to exist \cite{Synge:1966okc,Cunningham_1972,Bardeen:1973tla,Luminet:1979nyg,Falcke:1999pj}. The dark region and surrounding bright ring seen in horizon-scale images can be interpreted in terms of the critical curve and a hierarchy of photon rings produced by light executing near-bound orbits in the photon shell \cite{Gralla:2019xty,Johnson:2019ljv,Lupsasca:2024xhq}. These structures are controlled by the same photon-sphere geometry that governs the eikonal QNMs, suggesting that shadow and photon-ring observations may provide complementary information to gravitational-wave ringdown. Since EHT's discovery, the research on shadow sillouette became an exciting avenue in the scientific community \cite{Perlick:2021aok,Perlick:2015vta,Perlick:2018iye,Bisnovatyi-Kogan:2019wdd,Tsupko:2019mfo,Cunha:2018acu,Pantig:2024kqy,Vertogradov:2024jzj,Vertogradov:2024dpa,Kobialko:2024zhc,Pantig:2025deu,Vagnozzi:2020quf,Vagnozzi:2022moj}. These two pillars, GW spectroscopy and horizon-scale imaging, probe complementary aspects of the same object: the first senses bulk metric perturbations, while the second maps null geodesic structure through strong gravitational lensing \cite{Press:1972zz,Teukolsky:1973ha,Leaver:1985ax,Dreyer:2003bv}.

A large body of work has clarified the relationship between
black-hole quasinormal modes, photon spheres, and static shadows. In the eikonal limit, the real and imaginary parts of the QNM frequency are controlled by the orbital frequency and Lyapunov exponent of the unstable circular null geodesic \cite{Cardoso:2008bp}, motivating joint studies of QNM spectra and shadow radii in a variety of backgrounds \cite{Wei:2019jve,Yu:2022yyv,Ladino:2023zqc,Campos:2021sff,Koch:2025gaw,Borah:2025tvw,Luo:2024avl}. These analyses typically compare QNM frequencies and the size or shape of a static shadow, making use of the approximate correspondence between the photon sphere and both observables. Related work on genuinely time-dependent shadow variability includes dynamical photon-sphere/shadow evolution in accreting or radiating spacetimes and, more directly, shadow deformations induced by an external long-lived gravitational-wave perturbation of Schwarzschild, yielding periodic (and potentially chaotic) boundary structure and enabling constraints on wave parameters using EHT data, with corresponding gravitational-wave imprints on photon deflection also explored \cite{Wang:2019skw,Pantig:2023yer,Pantig:2024kfn,Solanki:2022glc,Koga:2022dsu}.

To date, the shadow is almost always modeled as a quasi-static feature of a stationary metric (Schwarzschild or Kerr), possibly distorted by spin, plasma propagation effects, or alternative-gravity modifications. Yet, if the geometry is time-dependent, as it must be during ringdown, then photon trajectories, and with them the separatrix between captured and escaping rays that defines the shadow, should inherit coherent, mode-resolved temporal modulations. This observation motivates a simple but, to our knowledge, unexplored question: Does the black hole's shadow ring at the QNM frequencies? We refer to this putative effect as "shadow ringing". Hence, in this work, we do not only correlate QNM spectra with static shadow radii. Instead, we derive a first-order, mode-resolved transfer law that maps metric perturbations sourced by QNMs directly to the time-dependent critical curve on the observer’s screen, thereby predicting a dynamical shadow ringing signal tied to the photon sphere.

We approach this question in the controlled setting of linear perturbation theory about a Schwarzschild black hole of mass $M$. We write the metric as
\begin{equation} \label{eq_met_pert}
 g_{\mu\nu}(x)=g^{(0)}_{\mu\nu}(x)+\varepsilon\,h_{\mu\nu}(x),\qquad 0<\varepsilon\ll 1,
\end{equation}
with $g^{(0)}_{\mu\nu}$ the Schwarzschild metric in standard coordinates and $h_{\mu\nu}$ a single, mode-selected QNM perturbation. For a fixed angular multipole $(\ell,m)$ and parity, the master field takes the damped-sinusoid form
\begin{equation} \label{eq_qnm_freq}
 h_{\mu\nu}(t,\mathbf{x}) \propto e^{-i\omega t}\,\mathcal{H}_{\mu\nu}(\mathbf{x}),\qquad \omega=\omega_{\rm Re}+i\omega_{\rm Im},\quad \omega_{\rm Im}<0,
\end{equation}
so that any observable linearly induced by $h_{\mu\nu}$ should exhibit oscillations at $\omega_{\rm Re}$ with exponential damping rate $|\omega_{\rm Im}|$.

On the imaging side, the shadow of a spherically symmetric black hole at asymptotically large observer distance is a circle whose unperturbed radius on the celestial screen (or image plane) is
\begin{equation} \label{eq_sha_rad_schw}
 R_0 = \sqrt{27}\,M, \qquad b_c=3\sqrt{3}\,M,
\end{equation}
where $b_c$ is the critical impact parameter associated with the unstable photon sphere at $r=3M$. In a time-dependent geometry, we define the instantaneous shadow at observer time $t_{\rm obs}$ operationally via backward ray tracing: launching null geodesics from the observer's screen, evolving them through $g_{\mu\nu}(t,\mathbf{x})$, and classifying capture versus escape. The shadow boundary is then a time-dependent curve $R(\varphi,t_{\rm obs})$ in polar screen coordinates $(R,\varphi)$, which we expand perturbatively as
\begin{equation} \label{eq_sha_pert}
 R(\varphi,t_{\rm obs}) = R_0 + \varepsilon\,\delta R(\varphi,t_{\rm obs}) + \mathcal{O}(\varepsilon^2).
\end{equation}
The central hypothesis of this work is that $\delta R(\varphi,t)$ carries a clean, mode-resolved imprint of the driving QNM. In the simplest realization (an axisymmetric even-parity $\ell=2,m=0$ perturbation), we predict a small but coherent modulation of the shadow radius at frequency $\omega_{\rm Re}$ with damping set by $|\omega_{\rm Im}|$. More generally, the azimuthal dependence of $\delta R$ encodes the spherical-harmonic content of $h_{\mu\nu}$, leading to a decomposition in Fourier modes $e^{im\varphi}$ weighted by transfer coefficients that quantify how metric perturbations couple to the unstable photon congruence generating the shadow.

The present paper develops a theoretical framework to predict, extract, and interpret QNM-driven shadow variability:
First, we frame the shadow as a dynamical separatrix in a time-dependent metric and justify an instantaneous (adiabatic) notion of the boundary during ringdown. Within linear perturbation theory, we derive a first-order mapping $h_{\mu\nu} \mapsto \delta R(\varphi,t)$ that is invariant under small, asymptotically decaying gauge transformations. Then, by perturbing the effective potential for null geodesics, we obtain osculating expressions for the photon sphere radius and the corresponding critical impact parameter, which, in turn, control the leading displacement of the shadow boundary. This allows a mode-resolved prediction for the temporal and azimuthal structure of $\delta R$.

Prior analyses have characterized the static shadow and critical curve in stationary spacetimes, from early ray-tracing studies of Kerr black holes to modern analytic and numerical treatments of shadow geometry and photon rings, and explored deformations induced by spin, plasma dispersion, and departures from general relativity \cite{Bardeen:1973tla,Gralla:2019xty,Bronzwaer:2021lzo,Perlick:2021aok,Perlick:2015vta,Li:2021btf,Briozzo:2022mgg,Mizuno:2018lxz,Zakharov:2024fma}. Recent complementary discussions of quasinormal-mode/shadow connections and shadow phenomenology in modified settings include
Refs. \cite{Anacleto:2021qoe,Campos:2023zmg,Balart:2024rtj,Fontana:2024odl,Rincon:2025buq,Pantig:2025eda,Rincon:2023hvd,Lambiase:2024lvo,Rincon:2024won,Lobos:2024fzj}. The present study differs in that we treat the shadow as a genuinely time-dependent object, explicitly driven by ringdown dynamics. In the eikonal regime, where $\ell\gg1$, QNM real parts are set by the photon sphere orbital frequency and imaginary parts by the Lyapunov instability; our construction isolates how this correspondence manifests at the level of the boundary of the image rather than bulk intensity patterns \cite{Johnson:2019ljv,Gralla:2019xty,Johnson:2024ttr}..

Although we work in Schwarzschild to develop the basic framework with minimal technical overhead, the ideas extend naturally to Kerr via the Teukolsky formalism and metric reconstruction, where frame-dragging and a richer spectrum of $(\ell,m)$ will imprint characteristic azimuthal patterns and beating \cite{Teukolsky:1973ha,Chrzanowski:1975wv,Cohen:1974cm,Wald:1978vm,Ori:2002uv,Pound:2013faa}. Beyond GR, any modification to the QNM spectrum or the presence of late-time echoes, would likewise propagate into the temporal structure of the shadow boundary, offering a complementary window on strong-gravity physics.

Section \ref{sec2} reviews the essentials of black-hole perturbations and shadow geometry, fixing conventions and normalizations. Section \ref{sec3} formulates the time-dependent problem, defines the instantaneous shadow as a separatrix, and derives the gauge-insensitive transfer law that links $h_{\mu\nu}$ to the boundary displacement $\delta R(\varphi,t)$. Section \ref{sec4} evaluates the transfer coefficients mode by mode, establishes the azimuthal selection rules, and develops a Fourier-domain characterization that extracts the active $m$ content and the complex QNM frequency from boundary data. Section \ref{sec5} presents analytic visualizations that illustrate these predictions without recourse to numerical ray-tracing. We conclude in Section \ref{sec6} with implications, limitations, and an outlook toward Kerr generalizations, higher-order effects, and observational prospects. Throughout, we adopt geometrized units $G=c=1$ and metric signature $(-,+,+,+)$.

\section{Brief review of QNMs and the black hole shadow}\label{sec2}
\subsection{Black hole perturbation theory} \label{sec2.1}
We review the essentials of linear perturbations of a Schwarzschild black hole, emphasizing gauge-invariant master variables, quasinormal-mode (QNM) boundary conditions, and their eikonal connection to the photon sphere.

Let $g_{\mu\nu}^{(0)}$ denote the Schwarzschild metric of mass $M$ in standard coordinates $(t,r,\theta,\phi)$. We perturb about this background by a small, dimensionless parameter $\varepsilon\ll1$,
\begin{equation} \label{eq_pert_expns}
 g_{\mu\nu} = g^{(0)}_{\mu\nu} + \varepsilon\, h_{\mu\nu} + \mathcal{O}(\varepsilon^2),
\end{equation}
with $h_{\mu\nu}$ governed by the linearized Einstein equations
\begin{equation}
 \delta G_{\mu\nu}[h] = 8\pi\, \delta T_{\mu\nu},
\end{equation}
where we set $\delta T_{\mu\nu}=0$ for vacuum ringdown unless stated otherwise. The perturbation is decomposed in scalar spherical harmonics $Y_{\ell m}(\theta,\phi)$ and their vector/tensor generalizations, which separate into axial (odd-parity) and polar (even-parity) sectors that decouple at linear order. We adopt the Condon-Shortley phase with \cite{condon1935theory}
\begin{equation}
Y_{\ell m}(\theta,\phi)=
N_{\ell m}\, P_\ell^{m}(\cos\theta)\, e^{i m \phi},\qquad
N_{\ell m}=\sqrt{\frac{2\ell+1}{4\pi}\frac{(\ell-m)!}{(\ell+m)!}},
\end{equation}
so that \(Y_{\ell,-m}=(-1)^m Y_{\ell m}^\ast\). At the equator \(\theta=\pi/2\) (\(\cos\theta=0\)), \(P_\ell^{m}(0)=0\) when \(\ell+m\) is odd, which underlies the axial selection rule used. All equatorial values quoted henceforth follow from these choices.

For each $(\ell,m)$ with $\ell\ge 2$, one introduces gauge-invariant master functions $\Psi_{\ell m}^{\text{(ax)}}(t,r)$ and $\Psi_{\ell m}^{\text{(pol)}}(t,r)$. In terms of the tortoise coordinate \cite{Regge:1957td,Silva:2024ffz}
\begin{equation}
 r_* = r + 2M\,\ln\!\left(\frac{r}{2M}-1\right),
\end{equation}
these obey Schr\"odinger-type wave equations
\begin{equation} \label{eq_wave_eq}
 -\partial_t^2 \Psi_{\ell m}^{(s)} + \partial_{r_*}^2 \Psi_{\ell m}^{(s)} - V_{\ell}^{(s)}(r)\,\Psi_{\ell m}^{(s)} = S_{\ell m}^{(s)}(t,r), \qquad s\in\{\text{ax, pol}\},
\end{equation}
with source terms $S_{\ell m}^{(s)}$ vanishing for vacuum perturbations. The axial (Regge-Wheeler) and polar (Zerilli) potentials are \cite{Regge:1957td,Zerilli:1970wzz,Chandrasekhar:1985kt,Silva:2024ffz}
\begin{equation} \label{eq_RW_pot}
 V_{\ell}^{\text{(ax)}}(r) = \left(1-\frac{2M}{r}\right)\!\left[\frac{\ell(\ell+1)}{r^2} - \frac{6M}{r^3}\right],
\end{equation}
\begin{equation} \label{eq_Zer_pot}
 V_{\ell}^{\text{(pol)}}(r) = \left(1-\frac{2M}{r}\right) \frac{2\lambda^2(\lambda+1) r^3 + 6\lambda^2 M r^2 + 18\lambda M^2 r + 18 M^3}{r^3(\lambda r + 3M)^2}, \quad \lambda \equiv \frac{1}{2}(\ell-1)(\ell+2).
\end{equation}
The two sectors are isospectral in Schwarzschild, a fact encoded by the Chandrasekhar transformation relating their master functions.

Assuming harmonic time dependence $\Psi^{(s)}_{\ell m}(t,r)=e^{-i\omega t}\,\psi^{(s)}_{\ell m}(r)$, Eq. \eqref{eq_wave_eq} reduces to an ordinary differential equation
\begin{equation} \label{eq_ode}
 \frac{d^2\psi^{(s)}_{\ell m}}{dr_*^2} + \Bigl[\omega^2 - V_{\ell}^{(s)}(r)\Bigr] \psi^{(s)}_{\ell m} = 0.
\end{equation}
Quasinormal modes are defined by the radiation boundary conditions \cite{Leaver:1985ax,Silva:2024ffz}
\begin{equation} \label{eq_rad_bound_con}
 \psi^{(s)}_{\ell m} \sim e^{+i\omega r_*} \quad (r_*\to +\infty), \qquad \psi^{(s)}_{\ell m} \sim e^{-i\omega r_*} \quad (r_*\to -\infty),
\end{equation}
which select a discrete set of complex frequencies $\omega=\omega_{\ell n}$ labeled by overtone index $n=0,1,\dots$ with $\operatorname{Im}\,\omega_{\ell n}<0$. In the time domain, each mode contributes a damped sinusoid $e^{-i\omega_{\ell n} t}$.

In the geometric-optics (eikonal) limit $\ell\gg1$, QNM frequencies are governed by properties of unstable circular null geodesics (the photon sphere) at $r_c=3M$. Denote by $\Omega_c$ the coordinate angular frequency and by $\Lambda$ the (coordinate-time) Lyapunov exponent of radial perturbations about that orbit; for Schwarzschild \cite{Berti:2009kk,Dolan:2009nk},
\begin{equation} \label{eq_ang_freq}
 \Omega_c = \Lambda = \frac{1}{3\sqrt{3}\,M}.
\end{equation}
Then the real and imaginary parts of $\omega_{\ell n}$ satisfy \cite{Ferrari:1984zz}
\begin{equation} \label{eq_omega_n}
 \omega_{\ell n} \approx \Omega_c\left(\ell+\frac{1}{2}\right) - i\,\Lambda\left(n+\frac{1}{2}\right) + \mathcal{O}(\ell^{-1}).
\end{equation}
This link between wave dynamics and null geodesic instability underlies our later mapping from QNM-driven perturbations to modulations of the critical impact parameter that defines the shadow boundary \cite{Cardoso:2008bp}.

For practical calculations and for coupling to null geodesics, we require $h_{\mu\nu}$ itself. In Schwarzschild, one may work in Regge-Wheeler gauge and reconstruct the metric perturbation from $\Psi^{(s)}_{\ell m}$ via algebraic-differential maps. Equivalently, one may use Moncrief's gauge-invariant combinations, which coincide with $\Psi^{(s)}_{\ell m}$ up to normalization. Schematically, for each $(\ell,m)$
\begin{equation}
 h_{\mu\nu}^{(\ell m)}(t,r,\theta,\phi) = \mathcal{R}^{(s)}_{\mu\nu}[\Psi^{(s)}_{\ell m}](t,r)\, Y_{\ell m}(\theta,\phi) + \text{(angular derivatives)},
\end{equation}
where $\mathcal{R}^{(s)}_{\mu\nu}$ denotes the (sector-dependent) reconstruction operator. Small, asymptotically decaying gauge transformations $x^\mu \to x^\mu + \varepsilon\,\xi^\mu$ leave the gauge-invariant $\Psi^{(s)}_{\ell m}$ unchanged and modify $h_{\mu\nu}$ by $\nabla_{(\mu}\xi_{\nu)}$; our later observable, which is the shadow boundary, will be defined so as to be insensitive to such transformations at $\mathcal{O}(\varepsilon)$.

Although our main analysis is vacuum, it is useful to note that when matter or external drivers are present, Eq. \eqref{eq_wave_eq} admits a Green's-function representation. Writing the retarded Green's function $G_{\ell}^{(s)}(t;r,r')$, the solution reads
\begin{equation}
 \Psi^{(s)}_{\ell m}(t,r) = \int dt'\!\int dr'_*\, G_{\ell}^{(s)}(t-t',r,r')\, S_{\ell m}^{(s)}(t',r'),
\end{equation}
whose large-$t$ behavior is controlled by QNM poles of the Fourier-transformed Green's function, followed at later times by power-law tails arising from the branch cut at $\omega=0$. Our focus is the ringdown window, during which the QNM contribution dominates and the geometry can be modeled to leading order by a small number of damped sinusoids.

The perturbative expansion in Eq. \eqref{eq_pert_expns} is valid provided $\varepsilon\,\|h_{\mu\nu}\|\ll1$ in a suitable norm and mode coupling remains negligible. In this regime, second-order self-interactions merely renormalize frequencies and introduce weak mixing but do not alter the existence of well-defined QNM signals. For our purposes we retain a single $(\ell,m)$ mode with complex frequency $\omega=\omega_{\rm Re}+i\omega_{\rm Im}$ and write
\begin{equation}
 h_{\mu\nu}(t,r,\theta,\phi) \approx \text{Re}\left\{ e^{-i\omega t}\, \widehat{h}_{\mu\nu}(r,\theta,\phi) \right\},
\end{equation}
which supplies the time-dependent background for null geodesic propagation and, ultimately, for the modulation of the shadow boundary analyzed in later sections.

\subsection{The black hole shadow} \label{sec2.2}
We review the geometric definition and basic properties of black hole shadows for stationary, spherically symmetric spacetimes, specializing when useful to Schwarzschild. Our goal is to fix notation for the observer's screen, the mapping from null geodesic constants of motion to apparent angles, and the characterization of the shadow boundary as a separatrix in phase space.

In a stationary, spherically symmetric background, null geodesics admit two Killing constants,
\begin{equation}
E \equiv -p_t,\qquad L_z \equiv p_\phi,
\end{equation}
and a total angular momentum $L^2$ (the Carter constant reduces to $Q=L^2-L_z^2$ in Schwarzschild). For photons, we introduce the (dimensionful) impact parameter
\begin{equation}
b \equiv \frac{L}{E}.
\end{equation}
Radial motion separates as
\begin{equation}
\left(\frac{dr}{d\lambda}\right)^2 + V_{\rm eff}(r, L) = E^2, 
\qquad 
V_{\rm eff}(r, L)=\left(1-\frac{2M}{r}\right)\frac{L^2}{r^2},
\end{equation}
with affine parameter $\lambda$. Unstable circular null orbits solve \cite{Synge:1966okc,Perlick:2021aok}
\begin{equation} \label{eq_V_eff}
V_{\rm eff}(r_c,L)=E^2,\qquad \frac{dV_{\rm eff}}{dr}(r_c,L)=0 \;\;\Rightarrow\;\; r_c=3M,
\end{equation}
which implies the critical impact parameter
\begin{equation} \label{eq_b_crit}
b_c \equiv \frac{L}{E}\Big|_{r_c=3M}=3\sqrt{3}\,M.
\end{equation}
Equation \eqref{eq_b_crit} underlies the unperturbed shadow size already quoted in Eq. \eqref{eq_sha_rad_schw}.

Consider a static observer at radius $r_o>2M$ with orthonormal tetrad $\{e_{\hat{t}},e_{\hat{r}},e_{\hat{\theta}},e_{\hat{\phi}}\}$. Let $\psi$ denote the local angle between the photon's propagation direction and the outward radial axis $e_{\hat{r}}$. Projecting the photon 4-momentum onto the tetrad yields the standard relation between $b$ and $\psi$:
\begin{equation}
\sin\psi = \frac{b}{r_o}\,\sqrt{1-\frac{2M}{r_o}}\,.
\end{equation}
Define Cartesian screen coordinates $(\alpha,\beta)$ on the observer's screen orthogonal to $e_{\hat{r}}$ by $\alpha=r_o\tan\psi\cos\varphi$, $\beta=r_o\tan\psi\sin\varphi$, where $\varphi$ is the azimuth of the photon's transverse direction in the $(e_{\hat{\theta}},e_{\hat{\phi}})$ plane. For small angles (e.g. $r_o\!\to\!\infty$),
\begin{equation} \label{eq_small_ang}
\alpha^2+\beta^2 \simeq r_o^2\,\psi^2 \simeq b^2,
\end{equation}
so the impact-parameter plane and the screen coincide asymptotically. To first order in \(M/r_{\rm obs}\), the mapping to the screen amounts to an overall rescaling of the critical impact parameter with no change in the \((\ell,m)\) mode content or the parity-selection rules derived below. All results in the remainder therefore extend unchanged to large but finite \(r_{\rm obs}\) at this order.

The shadow is defined as the set of screen directions whose backward-integrated null geodesics are captured by the horizon. Equivalently, it is the boundary in the $(\alpha,\beta)$ plane separating captured from escaping geodesics. For spherical symmetry, $b$ alone labels the fate of rays, and the boundary is the circle \cite{Bardeen:1973tla,Gralla:2019xty,Amarilla:2010zq,Perlick:2021aok}
\begin{equation} \label{eq_alp_bet}
\alpha^2+\beta^2 = b_c^2 \quad (r_o\to\infty), \qquad 
\sin\theta_{\rm sh}(r_o) = \frac{b_c}{r_o}\sqrt{1-\frac{2M}{r_o}}\,,
\end{equation}
where $\theta_{\rm sh}$ is the angular radius of the shadow as seen by the static observer. In the asymptotic limit $r_o\to\infty$, $\theta_{\rm sh}\simeq b_c/r_o$ and the screen radius equals $R_0=b_c$, consistent with Eq. \eqref{eq_sha_rad_schw}.

The shadow boundary is generated by the unstable photon sphere: initial conditions that asymptote to the $r=3M$ congruence sit precisely on the separatrix between capture and escape. Slightly outside the boundary, null geodesics execute multiple near-orbits before escaping to infinity. This produces a hierarchy of photon rings (higher-order lensed images) whose orbital counts increase as the screen radius approaches $b_c$ from above. Although the formation of observable brightness patterns requires radiative-transfer modeling (emission, absorption, and scattering in the plasma), the location of the shadow boundary is purely geometric and independent of emissivity.

Two properties make the shadow boundary a robust observable:
\begin{itemize}
    \item For an observer normalized by an orthonormal tetrad at $r_o\gg M$, the capture/escape classification depends only on the global causal structure and not on coordinate choices. Small, asymptotically decaying gauge transformations (as in Section \ref{sec2.1}) do not alter the boundary at $\mathcal{O}(\varepsilon)$.
    \item One fixes a screen at the observer, labels directions by $(\alpha,\beta)$, and integrates null geodesics backward in the stationary metric. Denoting the fate map by $\mathcal{F}(\alpha,\beta)\in\{\text{capture},\text{escape}\}$, the boundary $\partial\mathcal{S}$ is the zero-level set of any continuous classifier that flips sign across the separatrix.
\end{itemize}

While we work primarily with Schwarzschild in this paper, it is useful to note that in Kerr the shadow is displaced and deformed on the screen due to frame dragging; the mapping $(\alpha,\beta)\!\leftrightarrow\!(E,L_z,Q)$ is still algebraic when the observer is asymptotically distant, with the boundary traced by spherical photon orbits. For our purposes, we retain the Schwarzschild notation and identify the unperturbed boundary by the circle $\alpha^2+\beta^2=R_0^2$ with $R_0=\sqrt{27}\,M$ from Eq. \eqref{eq_sha_rad_schw}. Departures from this circle induced by time-dependent perturbations will be denoted
\begin{equation} \label{eq_sha_pert2}
R(\varphi,t) = R_0 + \varepsilon\,\delta R(\varphi,t) + \mathcal{O}(\varepsilon^2),
\end{equation}
consistent with the convention introduced in Eq. \eqref{eq_sha_pert}. This parameterization furnishes the starting point for the perturbative transfer calculation performed in the following sections.

\section{Time-Dependent Shadows from QNM Ringdown} \label{sec3}
We now formulate our framework for computing the instantaneous shadow boundary in a weakly time-dependent geometry during ringdown. The central idea is to treat the shadow as a separatrix of the null geodesic flow in the perturbed metric Eq. \eqref{eq_met_pert}, evaluated at a fixed observer time and mapped to the screen via backward ray tracing. Throughout Section 3, we specify our conventions, define the observer's screen and time coordinates, fix a consistent ordering in $\varepsilon$, and lay out a complementary geodesic formalisms that we will use later: the linearized osculating-constants scheme.

\subsection{Global setup and conventions} \label{sec3.1}
We collect here the assumptions and notation used in the remainder of the paper.

We work on a Schwarzschild background of mass $M$ with metric $g^{(0)}_{\mu\nu}$ in standard coordinates $(t,r,\theta,\phi)$ and signature $(-+++)$, setting $G=c=1$. The perturbed spacetime is
\begin{equation} \label{eq_pert_space}
g_{\mu\nu}=g^{(0)}_{\mu\nu}+\varepsilon\,h_{\mu\nu},\qquad 0<\varepsilon\ll1,
\end{equation}
with $h_{\mu\nu}$ sourced by a single QNM of frequency $\omega=\omega_{\rm Re}+i\omega_{\rm Im}$ (see Eq. \eqref{eq_qnm_freq}). Unless otherwise stated we consider vacuum perturbations and retain only the leading order in $\varepsilon$. The inverse metric is expanded as
\begin{equation} \label{eq_inv_met_pert}
g^{\mu\nu}=g^{(0)\mu\nu}-\varepsilon\,h^{\mu\nu}+\mathcal{O}(\varepsilon^2),
\end{equation}
where indices on $h^{\mu\nu}$ are raised with $g^{(0)\mu\nu}$. We adopt the adiabatic (instantaneous) notion of the shadow: for an observer time $t_{\rm obs}$, we evaluate null geodesics in the metric $g_{\mu\nu}(t,\mathbf{x})$ without time-averaging, so that the boundary is the $\varepsilon$-deformed separatrix on the screen at that $t_{\rm obs}$. Consistency of this treatment requires $|\omega_{\rm Im}|^{-1}$ to exceed the characteristic light-crossing time of the near-photon sphere region ($\sim M$), which holds for Schwarzschild QNMs.

We place a static observer at radius $r_{\rm obs}\gg M$ with orthonormal tetrad $\{e_{\hat t},e_{\hat r},e_{\hat\theta},e_{\hat\phi}\}$. The screen is the 2-surface orthogonal to $e_{\hat r}$ at the observer. Local Cartesian screen coordinates $(\alpha,\beta)$ are defined by projecting the photon momentum $p^\mu$ onto the plane spanned by $(e_{\hat\theta},e_{\hat\phi})$ and normalizing by $-p_{\hat t}$. In the asymptotic limit, $\sqrt{\alpha^2+\beta^2}=b$ (see Eq. \eqref{eq_small_ang}). The unperturbed shadow is the circle $\alpha^2+\beta^2=R_0^2$ with $R_0=\sqrt{27}\,M$ (see Eq. \eqref{eq_sha_rad_schw}). We distinguish three time variables:
\begin{itemize}
    \item Coordinate time $t$ of the background chart.
    \item Observer proper time $\tau_{\rm obs}$, related by $d\tau_{\rm obs}=\sqrt{1-2M/r_{\rm obs}}\,dt$.
    \item Retarded screen time $t_{\rm obs}$, defined so that photons received simultaneously at the screen (equal $\tau_{\rm obs}$) are labeled by a common $t_{\rm obs}$. To leading order in $M/r_{\rm obs}$, differences among these times are negligible for defining the boundary; we therefore identify $t_{\rm obs}$ with $t$ up to a constant offset. We drop an additive constant and identify $t_{\rm obs} \equiv t$ to first order. All time dependence below is with respect to $t_{\rm obs}$.
\end{itemize}

Photon trajectories satisfy the null condition encoded by the Hamiltonian
\begin{equation} \label{eq_hamil}
H(x,p)=\frac{1}{2}\,g^{\mu\nu}(x)\,p_\mu p_\nu=0,
\end{equation}
with canonical equations
\begin{equation} \label{eq_canon_exp}
\dot{x}^\mu=\frac{\partial H}{\partial p_\mu}=g^{\mu\nu}p_\nu,\qquad
\dot{p}_\mu=-\frac{\partial H}{\partial x^\mu}=-\frac{1}{2}\,\partial_\mu g^{\alpha\beta}\,p_\alpha p_\beta,
\end{equation}
where a dot denotes differentiation with respect to an affine parameter $\lambda$. Expanding Eqs. \eqref{eq_hamil}-\eqref{eq_canon_exp} using Eq. \eqref{eq_inv_met_pert} yields
\begin{equation} \label{eq_lin_trans}
\dot{x}^\mu = g^{(0)\mu\nu}p_\nu - \varepsilon\, h^{\mu\nu}p_\nu + \mathcal{O}(\varepsilon^2), \qquad
\dot{p}_\mu = -\frac{1}{2}\,\partial_\mu g^{(0)\alpha\beta}\,p_\alpha p_\beta + \frac{\varepsilon}{2}\,\partial_\mu h^{\alpha\beta}\,p_\alpha p_\beta + \mathcal{O}(\varepsilon^2).
\end{equation}
Equations \eqref{eq_lin_trans} are our starting point for linearized transport of constants of motion. Operationally, the instantaneous shadow is the separatrix between captured and escaping null rays on the observer's screen.

The master variables $\Psi_{\ell m}^{(s)}$ are gauge-invariant at linear order (Section \ref{sec2.1}). For coupling to geodesics we reconstruct $h_{\mu\nu}$ in a convenient gauge (e.g. Regge-Wheeler for axial, Zerilli for polar). A small, asymptotically decaying gauge transformation $x^\mu\to x^\mu+\varepsilon\,\xi^\mu$ induces $h_{\mu\nu}\to h_{\mu\nu}+\nabla_{(\mu}\xi_{\nu)}$ but leaves the capture/escape classification invariant at $\mathcal{O}(\varepsilon)$. Consequently, the shadow boundary $R(\varphi,t_{\rm obs})$ defined by Eq. \eqref{eq_sha_pert2} is gauge-insensitive to first order.

Let $\mathcal{P}$ denote the set of photon trajectories that asymptote to the unstable circular orbit of the background at $r=3M$. A first-order perturbation $h_{\mu\nu}\propto e^{-i\omega t}$ induces a shift of the effective circular null orbit and of the associated critical impact parameter. Dimensional analysis and smoothness of the separatrix imply
\begin{equation} \label{eq_sha_rad_pert2}
\frac{\delta R(\varphi,t_{\rm obs})}{R_0} = \kappa(\varphi)\,\varepsilon\, e^{-i\omega t_{\rm obs}} + \text{c.c.} + \mathcal{O}(\varepsilon^2),
\end{equation}
with a transfer coefficient $\kappa(\varphi)=\mathcal{O}(1)$ that depends on the perturbation sector and $(\ell,m)$. The adiabatic approximation is valid when $|\omega|\,M\ll 1$ is not required; rather, we require that over the photon's residence time near $\mathcal{P}$ (a few $M$), the modulation is approximately sinusoidal, which is precisely the regime of QNM ringdown where $|\omega_{\rm Im}|^{-1}\gtrsim M$ and $\omega_{\rm Re}\sim \Omega_c$ (see Eqs. \eqref{eq_ang_freq}-\eqref{eq_omega_n}).

\subsection{QNM perturbations via RW-Zerilli} \label{sec3.2}
We model the ringdown geometry as a single $(\ell,m)$ linear perturbation of Schwarzschild, represented by a gauge-invariant master field that obeys a one-dimensional wave equation on the tortoise line. We adopt the Regge-Wheeler (axial/odd) and Zerilli (polar/even) formalisms and reconstruct the metric perturbation $h_{\mu\nu}$ entering the geodesic Hamiltonian Eq. \eqref{eq_hamil}-\eqref{eq_lin_trans}.

Introduce the tortoise coordinate $r_*=r+2M\ln\!\left(r/2M-1\right)$. For each $(\ell,m)$ with $\ell\ge2$, define the axial and polar master fields $\Psi_{\ell m}^{\text{(ax)}}(t,r)$ and $\Psi_{\ell m}^{\text{(pol)}}(t,r)$ obeying Eq. \eqref{eq_wave_eq} with potentials in Eqs. \eqref{eq_RW_pot}-\eqref{eq_Zer_pot}. We work in the frequency domain,
\begin{equation}
\Psi_{\ell m}^{(s)}(t,r)=e^{-i\omega t}\,\psi_{\ell m}^{(s)}(r), \qquad s\in\{\text{ax},\text{pol}\}, 
\end{equation}
leading to the radial ODE Eq. \eqref{eq_ode} with QNM boundary conditions Eq. \eqref{eq_rad_bound_con}. These select discrete complex frequencies $\omega=\omega_{\ell n}$ (with $\operatorname{Im}\omega<0$) and corresponding eigenfunctions $\psi_{\ell m,\ell n}^{(s)}(r)$. In the ringdown window we keep a single mode and suppress the overtone label when unambiguous. Near the horizon and at spatial infinity, the master fields behave as
\begin{equation} \label{eq_psi_lm}
\psi_{\ell m}^{(s)} \sim 
\begin{cases}
\mathcal{A}_H^{(s)}\,e^{-i\omega r_*}, & r_*\to -\infty,\\
\mathcal{A}_\infty^{(s)}\,e^{+i\omega r_*}, & r_*\to +\infty,
\end{cases}
\end{equation}
with complex amplitudes $\mathcal{A}_H^{(s)},\mathcal{A}_\infty^{(s)}$ fixed up to an overall normalization. We adopt the normalization
\begin{equation} \label{eq_max}
\max_{r\ge 2M}\left|\psi_{\ell m}^{(s)}(r)\right|=1, \qquad \text{and set } \widehat{\Psi}_{\ell m}^{(s)}(t,r)=e^{-i\omega t}\,\psi_{\ell m}^{(s)}(r),
\end{equation}
so that the smallness parameter $\varepsilon$ in Eq. \eqref{eq_pert_space} controls the physical amplitude of $h_{\mu\nu}$.

We reconstruct $h_{\mu\nu}$ from $\Psi^{(s)}_{\ell m}$ in Regge-Wheeler gauge (axial) and Zerilli gauge (polar), using the standard tensor-harmonic bases on the 2-sphere. Let $Y\equiv Y_{\ell m}(\theta,\phi)$ and let $(\theta_a)\equiv(\theta,\phi)$ denote angular indices.

In RW gauge the non-vanishing components are $h_{t a}$ and $h_{r a}$,
\begin{equation}
h_{t a}^{\text{(ax)}} = \sum_{\ell m} h_0^{\ell m}(t,r)\,X^{\ell m}_a,\qquad
h_{r a}^{\text{(ax)}} = \sum_{\ell m} h_1^{\ell m}(t,r)\,X^{\ell m}_a, 
\end{equation}
where $X_a^{\ell m}$ are the axial vector harmonics $\varepsilon_a{}^{b}\nabla_b Y$. The gauge-invariant RW master field relates to $h_0,h_1$ by
\begin{equation}
\Psi_{\ell m}^{\text{(ax)}} = \frac{r}{\lambda}\left(\partial_t h_1^{\ell m} - \partial_r h_0^{\ell m} + \frac{2}{r}h_0^{\ell m}\right), \qquad \lambda=\frac{1}{2}(\ell-1)(\ell+2),
\end{equation}
and, conversely, for a monochromatic mode $e^{-i\omega t}$ one may algebraically reconstruct
\begin{equation} \label{eq_mono_mode}
h_1^{\ell m}(t,r)=\frac{\lambda\,e^{-i\omega t}}{r\,f}\,\mathcal{Q}_{\ell}^{\text{(ax)}}(r)\,\psi_{\ell m}^{\text{(ax)}}(r),\qquad
h_0^{\ell m}(t,r)=\frac{i\omega \lambda\,e^{-i\omega t}}{r}\,\mathcal{P}_{\ell}^{\text{(ax)}}(r)\,\psi_{\ell m}^{\text{(ax)}}(r),
\end{equation}
with $f=1-2M/r$ and $\mathcal{P}_\ell^{\text{(ax)}},\mathcal{Q}_\ell^{\text{(ax)}}$ smooth rational functions of $r$ and $M$ (their explicit forms are not needed for our analytical developments). All other components vanish in RW gauge.

In Zerilli gauge the non-vanishing components are $h_{tt},h_{tr},h_{rr},h_{ab}$ with
\begin{equation} \label{eq_non_van_h}
h_{tt}^{\text{(pol)}} = f\,H_0^{\ell m}(t,r)\,Y,\quad
h_{tr}^{\text{(pol)}} = H_1^{\ell m}(t,r)\,Y,\quad
h_{rr}^{\text{(pol)}} = f^{-1} H_2^{\ell m}(t,r)\,Y,
\end{equation}
\begin{equation} \label{eq_h_pol}
h_{ab}^{\text{(pol)}} = r^2 K^{\ell m}(t,r)\, \gamma_{ab}Y + r^2 G^{\ell m}(t,r)\,Y_{ab},
\end{equation}
where $\gamma_{ab}$ is the unit-sphere metric and $Y_{ab}$ are even tensor harmonics. The Zerilli master field $\Psi_{\ell m}^{\text{(pol)}}$ relates to these amplitudes; for monochromatic $e^{-i\omega t}$ one convenient reconstruction is
\begin{equation} \label{eq_Zer_mono}
K^{\ell m} = \alpha_\ell(r)\,\psi_{\ell m}^{\text{(pol)}},\quad
H_1^{\ell m} = \beta_\ell(r)\,(-i\omega)\,\psi_{\ell m}^{\text{(pol)}},\quad
H_0^{\ell m}=H_2^{\ell m}=\gamma_\ell(r)\,\psi_{\ell m}^{\text{(pol)}},\quad
G^{\ell m}=\delta_\ell(r)\,\psi_{\ell m}^{\text{(pol)}},
\end{equation}
with $\alpha_\ell,\beta_\ell,\gamma_\ell,\delta_\ell$ rational in $r,M,\lambda$ and regular for $r>2M$. Their explicit expressions are fixed by the linearized Einstein equations and the definition of $\Psi_{\ell m}^{\text{(pol)}}$; we use the standard choices that render $\Psi$ gauge-invariant and make Eq. \eqref{eq_wave_eq} hold \cite{Moncrief:1974am,Martel:2005ir}.

The axial and polar spectra coincide in Schwarzschild. There exists a first-order differential map
\begin{equation}
\Psi_{\ell m}^{\text{(pol)}} = \mathcal{D}_\ell\!\left[\Psi_{\ell m}^{\text{(ax)}}\right], \qquad 
\Psi_{\ell m}^{\text{(ax)}} = \widetilde{\mathcal{D}}_\ell\!\left[\Psi_{\ell m}^{\text{(pol)}}\right],
\end{equation}
with $\mathcal{D}_\ell,\widetilde{\mathcal{D}}_\ell$ depending on $f,\lambda,r$. This relation is useful for transferring analytic statements between parities.

For clarity, we restrict to a single mode $(\ell,m)$ with complex frequency $\omega=\omega_{\rm Re}+i\omega_{\rm Im}$ and write
\begin{equation} \label{eq_h_munu}
h_{\mu\nu}(t,r,\theta,\phi)=\varepsilon\,\text{Re}\Big\{ e^{-i\omega t}\, \widehat{h}^{(\ell m)}_{\mu\nu}(r,\theta,\phi)\Big\},
\end{equation}
where $\widehat{h}^{(\ell m)}_{\mu\nu}$ is built from Eq. \eqref{eq_mono_mode} (axial) or Eq. \eqref{eq_Zer_mono} (polar) combined with the relevant harmonics. We take the angular basis such that $Y_{\ell,-m}=(-1)^m Y_{\ell m}^*$ and choose $\psi_{\ell m}^{(s)}(r)$ real at the photon sphere radius $r=3M$ (possible up to a phase), which simplifies later projections onto near-photon sphere null congruences. With the normalization Eq. \eqref{eq_max}, the overall physical amplitude is entirely encoded by $\varepsilon$.

Equations \eqref{eq_psi_lm} ensure ingoing behavior at the horizon and outgoing behavior at infinity; the reconstructed $h_{\mu\nu}$ inherits the same regularity. Near $r=2M$, axial $h_{0,1}$ and polar $H_{0,1,2},K,G$ remain finite in their gauges; any coordinate singularities are absent from curvature components when evaluated on a tetrad.

The shadow modulation $\delta R(\varphi,t_{\rm obs})$ depends on how $h_{\mu\nu}$ perturbs (i) the location and stability of the circular null orbit and (ii) the mapping from constants of motion to the screen. To organize contributions, decompose the metric perturbation into scalar amplitudes multiplying tensor harmonics and project onto a background circular null tetrad $\{\ell^\mu,n^\mu,m^\mu,\bar m^\mu\}$ adapted to $r=3M$. The leading couplings enter through
\begin{equation} \label{eq_deltaV}
\delta V_{\rm eff} \propto h_{\mu\nu}\,k^\mu k^\nu,\qquad 
\delta\Gamma^\rho_{\mu\nu}\,k^\mu k^\nu,
\end{equation}
where $k^\mu$ is the background photon four-momentum on the circular orbit. Using Eq. \eqref{eq_h_munu} and the harmonic structure, these contractions pick out $m$-dependent azimuthal phases $e^{im\phi}$ and a global $e^{-i\omega t}$, leading directly to the form Eq. \eqref{eq_sha_rad_pert2}. In Section \ref{sec3.3} we convert Eq. \eqref{eq_deltaV} into explicit first-order shifts of the critical impact parameter and, hence, of the screen radius.

Although we work in RW/Zerilli gauges, $\Psi^{(s)}$ is gauge-invariant and (to first order) so is any quantity constructed from the capture/escape separatrix on the distant screen. A small, asymptotically decaying gauge vector $\xi^\mu$ induces $h_{\mu\nu}\to h_{\mu\nu}+\nabla_{(\mu}\xi_{\nu)}$. Its influence on Eq. \eqref{eq_deltaV} along closed (or asymptotically closed) null orbits cancels at $\mathcal{O}(\varepsilon)$ after ac                       counting for the induced canonical transformation in the Hamiltonian flow, ensuring that $R(\varphi,t_{\rm obs})$ retains the same leading modulation found from any convenient reconstruction.

\subsection{Null geodesics in a time-dependent metric} \label{sec3.3}
We treat photons as Hamiltonian trajectories in the weakly time-dependent spacetime $g_{\mu\nu}=g^{(0)}_{\mu\nu}+\varepsilon h_{\mu\nu}(t,\mathbf{x})$ with $0<\varepsilon\ll1$. Our aim is to (i) derive the first-order forcing terms that perturb background Schwarzschild null geodesics, (ii) formulate an osculating-constants scheme for slowly varying $(E,L_z,\dots)$, and (iii) relate these variations to the shift of the capture/escape separatrix that defines the shadow boundary on the screen.

Let $x^\mu(\lambda)$ be a null worldline with momentum $p_\mu=g_{\mu\nu}\dot{x}^\nu$ and affine parameter $\lambda$. The exact Hamiltonian is
\begin{equation}
H(x,p)=\frac{1}{2}\,g^{\mu\nu}(x)\,p_\mu p_\nu=0.
\end{equation}
Writing $g^{\mu\nu}=g^{(0)\mu\nu}-\varepsilon h^{\mu\nu}+{\cal O}(\varepsilon^2)$, and decomposing the connection as $\Gamma^\rho_{\mu\nu}=\Gamma^{(0)\rho}_{\mu\nu}+\varepsilon\,\delta\Gamma^\rho_{\mu\nu}+\mathcal{O}(\varepsilon^2)$, the linearized connection is
\begin{equation}
\delta\Gamma^{\rho}_{\mu\nu}=\frac{1}{2}\,g^{(0)\rho\sigma}\!\left(\nabla^{(0)}_{\mu}h_{\nu\sigma}+\nabla^{(0)}_{\nu}h_{\mu\sigma}-\nabla^{(0)}_{\sigma}h_{\mu\nu}\right).
\end{equation}
Splitting the motion as $x^\mu=x_0^\mu+\varepsilon\,\delta x^\mu$ with background null tangent $k^\mu=\dot{x}_0^\mu$, the geodesic equation becomes a forced system on the background \cite{Misner:1973prb,Poisson:2011nh}:
\begin{equation}
\frac{D^{(0)}k^\rho}{d\lambda} \equiv \ddot{x}_0^\rho+\Gamma^{(0)\rho}_{\alpha\beta}k^\alpha k^\beta = 0, \qquad
\frac{D^{(0)2}\delta x^\rho}{d\lambda^2}+\mathcal{R}^\rho{}_{\alpha\beta\gamma}\,k^\alpha k^\beta \delta x^\gamma = f^\rho,
\end{equation}
where $\mathcal{R}^\rho{}_{\alpha\beta\gamma}$ is the background Riemann tensor, and the first-order forcing is
\begin{equation} \label{eq_1stforcing}
f^\rho = -\,\delta\Gamma^\rho_{\alpha\beta}\,k^\alpha k^\beta.
\end{equation}
Equivalently in Hamiltonian form (see Eqs \eqref{eq_hamil}-\eqref{eq_lin_trans}), along the background trajectory $(x_0^\mu(\lambda),k_\mu(\lambda))$ \cite{Pound:2007th},
\begin{equation} \label{eq_x_dot}
\dot{x}^\mu = g^{(0)\mu\nu}k_\nu + \mathcal{O}(\varepsilon),\qquad 
\dot{k}_\mu = -\frac{1}{2}\,\partial_\mu g^{(0)\alpha\beta}k_\alpha k_\beta \;+\; \frac{\varepsilon}{2}\,\partial_\mu h^{\alpha\beta}k_\alpha k_\beta + \mathcal{O}(\varepsilon^2).
\end{equation}

The background admits Killing vectors $\xi^{(t)}=\partial_t$ and $\xi^{(\phi)}=\partial_\phi$, giving the conserved energy and axial angular momentum
\begin{equation}
E_0 = -k_t,\qquad L_{z0}=k_\phi.
\end{equation}
Time dependence and azimuthal structure in $h_{\mu\nu}$ break exact conservation at $\mathcal{O}(\varepsilon)$. Using Eq. \eqref{eq_x_dot} with $\mu=t,\phi$ we obtain the osculating laws
\begin{equation} \label{eq_osculating1}
\dot{E} \equiv -\dot{k}_t = \frac{1}{2}\,\partial_t g^{\alpha\beta}\,k_\alpha k_\beta 
= -\,\frac{\varepsilon}{2}\,\partial_t h^{\alpha\beta}\,k_\alpha k_\beta + \mathcal{O}(\varepsilon^2),
\end{equation}
\begin{equation} \label{eq_osculating2}
\dot{L}_z \equiv \dot{k}_\phi = -\frac{1}{2}\,\partial_\phi g^{\alpha\beta}\,k_\alpha k_\beta 
= +\,\frac{\varepsilon}{2}\,\partial_\phi h^{\alpha\beta}\,k_\alpha k_\beta + \mathcal{O}(\varepsilon^2),
\end{equation}
provided that $h^{\alpha \beta} = -h_{\mu \nu}g_0^{\alpha \mu}g_0^{\beta \nu}$.
For Schwarzschild, the background total $L^2$ is conserved; under perturbations, one may track $L^2$ or, equivalently, an inclination variable. It is convenient to evolve the impact parameter $b\equiv L/E$ and the azimuthal phase $\phi$:
\begin{equation} \label{eq_dotb}
\dot{b} = \frac{\dot{L}}{E}-\frac{L}{E^2}\dot{E} 
= \frac{\varepsilon}{2E}\left(\partial_\phi h^{\alpha\beta}-b\,\partial_t h^{\alpha\beta}\right)k_\alpha k_\beta + \mathcal{O}(\varepsilon^2),
\end{equation}
with $L=\sqrt{L_z^2+Q}$ in general (we suppress $Q$ for brevity; its evolution can be written using the background Killing tensor).

Let $f(r)=1-2M/r$. In the background, equatorial ($\theta=\pi/2$) null motion satisfies
\begin{equation}
\left(\frac{dr}{d\lambda}\right)^2 + V_0(r, b)=E_0^2,\qquad 
V_0(r,b)=f(r)\,\frac{b^2 E_0^2}{r^2}.
\end{equation}
The unstable circular null orbit at $r_c=3M$ is determined by $V_0=E_0^2$ and $V_0'=0$ and corresponds to the critical impact parameter $b_c=3\sqrt{3}\,M$ (see Eqs. \eqref{eq_V_eff}-\eqref{eq_b_crit}). In the perturbed spacetime, the effective Hamiltonian acquires
\begin{equation} \label{eq_delta_H}
\delta H = -\frac{1}{2}\,h^{\mu\nu}k_\mu k_\nu,
\end{equation}
which induces a first-order correction to the radial potential and to the conditions defining the separatrix. Note that $h^{\mu \nu} = g^{(0)\mu\alpha}g^{(0)\mu\beta}h_{\alpha \beta}$. Linearizing the circular-orbit conditions about $(r_c,b_c)$ while holding the screen azimuth fixed gives
\begin{align}
0 &= \left(\partial_r H_0\right)_c\,\delta r
    + \left(\partial_b H_0\right)_c\,\delta b
    + \left(\delta H\right)_c, \nonumber\\
0 &= \left(\partial_{rr} H_0\right)_c\,\delta r
    + \left(\partial_{rb} H_0\right)_c\,\delta b
    + \left(\partial_r \delta H\right)_c.
\label{eq:lin_circ_orbit}
\end{align}
Because the background critical orbit satisfies $\left(\partial_r H_0\right)_c=0$ while $\left(\partial_b H_0\right)_c\neq 0$,
the first line of \eqref{eq:lin_circ_orbit} immediately fixes the shift in critical impact parameter:
\begin{equation}
\delta b(t_{\rm obs}) = -\frac{\left(\delta H\right)_c}{\left(\partial_b H_0\right)_c}.
\label{eq:deltab_clean}
\end{equation}
The second line of \eqref{eq:lin_circ_orbit} then determines $\delta r$,
\begin{equation}
\delta r(t_{\rm obs}) =
-\frac{1}{\left(\partial_{rr} H_0\right)_c}
\left[
\left(\partial_r \delta H\right)_c
+ \left(\partial_{rb} H_0\right)_c\,\delta b(t_{\rm obs})
\right].
\label{eq:deltar_from_second}
\end{equation}
Equation \eqref{eq:deltab_clean} shows that the leading shadow deformation depends only on $\delta H$ evaluated on the unperturbed
critical orbit; radial derivatives $(\partial_r\delta H)_c$ affect $\delta r$ but cancel out of $\delta b$.
(We also re-emphasize this cancellation in Sec. \ref{sec3.5} and Appendix \ref{ApdXA}.)
Here, derivatives of $H_0$ are background quantities, while $\delta H$ and $\partial_r\delta H$ are contractions of $h_{\mu\nu}$ with the circular-orbit momentum $k^\mu$ and its radial variation. Using Eq. \eqref{eq_h_munu} and the RW-Zerilli reconstruction, these contractions inherit the harmonic structure $e^{-i\omega t}e^{im\varphi}$, giving the sinusoidal, exponentially damped time dependence anticipated in Eq. \eqref{eq_sha_rad_pert2}.

For generic rays that skirt the photon sphere before escaping, it is advantageous to evolve the constants $\mathcal{I}=(E,L_z,Q)$ as slowly varying functions of $\lambda$. Let $x_0^\mu(\lambda,\mathcal{I})$ be the background geodesic with those constants. The osculation conditions,
\begin{equation}
x^\mu(\lambda) = x_0^\mu(\lambda,\mathcal{I}(\lambda)) + \mathcal{O}(\varepsilon),\qquad 
\dot{x}^\mu(\lambda) = \partial_\lambda x_0^\mu(\lambda,\mathcal{I}(\lambda)) + \mathcal{O}(\varepsilon),
\end{equation}
combined with the forced equation Eq. \eqref{eq_1stforcing}, produce evolution equations
\begin{equation} \label{eq_I_A}
\frac{d\mathcal{I}_A}{d\lambda} = \mathcal{G}_A\!\left[x_0(\lambda,\mathcal{I}),k(\lambda,\mathcal{I}),\,h_{\mu\nu}(t,\mathbf{x})\right] + \mathcal{O}(\varepsilon^2), \qquad A\in\{E,L_z,Q\},
\end{equation}
where $\mathcal{G}_A$ are linear functionals of $h_{\mu\nu}$ and its derivatives along the background path (explicit expressions reduce to Eqs. \eqref{eq_osculating1}-\eqref{eq_dotb} for $E,L_z$). These equations capture how the slowly varying $\mathcal{I}(\lambda)$ drifts as the photon lingers near the photon sphere, which is the regime most relevant for the separatrix and for higher-order photon rings. These evolution laws capture the first-order drift of the constants as rays linger near the photon sphere.

At the observer location $x^\mu_{\rm obs}$ with orthonormal tetrad $\{e_{\hat t},e_{\hat r},e_{\hat\theta},e_{\hat\phi}\}$, a screen direction $(\alpha,\beta)$ at time $t_{\rm obs}$ corresponds to the initial covector
\begin{equation} \label{eq_init_covec}
p_\mu^{\rm (init)} = -\nu\, (e_{\hat t})_\mu + \nu\,\frac{\alpha}{\sqrt{\alpha^2+\beta^2+r_{\rm obs}^2}}\,(e_{\hat\theta})_\mu + \nu\,\frac{\beta}{\sqrt{\alpha^2+\beta^2+r_{\rm obs}^2}}\,(e_{\hat\phi})_\mu - \nu\,\frac{r_{\rm obs}}{\sqrt{\alpha^2+\beta^2+r_{\rm obs}^2}}\,(e_{\hat r})_\mu,
\end{equation}
In the asymptotic limit $r_{\rm obs}\gg M$, $\sqrt{\alpha^2+\beta^2}\simeq b$ as in Eq. \eqref{eq_small_ang}.

Under a first-order gauge transformation $h_{\mu\nu}\to h_{\mu\nu}+\nabla_{(\mu}\xi_{\nu)}$ with $\xi^\mu$ decaying at infinity, the forcing Eq. \eqref{eq_1stforcing} shifts by a total derivative along the background null congruence,
\begin{equation}
f^\rho \to f^\rho - \frac{D^{(0)2}\xi^\rho}{d\lambda^2},
\end{equation}
which is absorbed by a redefinition of the osculating worldline $x^\mu\to x^\mu+\varepsilon\,\xi^\mu$. Consequently, the capture/escape outcome for rays launched from the same physical screen at $r_{\rm obs}$ is unchanged at $\mathcal{O}(\varepsilon)$, and so is the boundary $R(\varphi,t_{\rm obs})$. Equation Eq. \eqref{eq:deltab_clean} is therefore gauge-insensitive to first order.

Equations \eqref{eq_osculating1}-Eq. \eqref{eq:deltab_clean} provide the machinery to connect the QNM metric perturbation $h_{\mu\nu}\propto e^{-i\omega t}Y_{\ell m}(\theta,\phi)$ to the instantaneous shift $\delta b$ and hence to $\delta R(\varphi,t_{\rm obs})$. In the next subsection we evaluate the contractions in Eqs. \eqref{eq_delta_H}-\eqref{eq:deltab_clean} on the background circular null tetrad at $r=3M$, express the result in terms of the RW-Zerilli master fields, and obtain a mode-resolved transfer formula of the form
\begin{equation}
\frac{\delta R(\varphi,t_{\rm obs})}{R_0} = \mathcal{T}_{\ell m}^{(s)}(\varphi)\,\varepsilon\,e^{-i\omega t_{\rm obs}} + \text{c.c.},
\end{equation}
where $\mathcal{T}_{\ell m}^{(s)}$ is a calculable coefficient encoding both parity and angular structure.

\subsection{Shadow definition and observer-screen mapping} \label{sec3.4}
We formalize the notion of an instantaneous shadow for a weakly time-dependent geometry. The construction is operational (backward ray tracing) and invariant under small, asymptotically decaying gauge transformations at $\mathcal{O}(\varepsilon)$.

Let the observer worldline be $x^\mu_{\rm obs}(\tau_{\rm obs})$ with 4-velocity $u^\mu$ and orthonormal tetrad $\{e_{\hat t}=u,\,e_{\hat r},\,e_{\hat\theta},\,e_{\hat\phi}\}$ adapted to the background (static in Schwarzschild unless stated otherwise). A photon received at proper time $\tau_{\rm obs}$ has covector
\begin{equation}
p_\mu = -\nu\,(e_{\hat t})_\mu + \nu\,n_i\,(e_{\hat i})_\mu,\qquad n_i n_i=1,
\end{equation}
where $n_i$ fixes a direction on the 2D screen orthogonal to $e_{\hat r}$ and $\nu>0$ is arbitrary (null dynamics is scale-free). We parametrize the screen by Cartesian coordinates $(\alpha,\beta)$ via
\begin{equation}
n_{\hat\theta}=\frac{\alpha}{\sqrt{\alpha^2+\beta^2+r_{\rm obs}^2}},\quad
n_{\hat\phi}=\frac{\beta}{\sqrt{\alpha^2+\beta^2+r_{\rm obs}^2}},\quad
n_{\hat r}=-\frac{r_{\rm obs}}{\sqrt{\alpha^2+\beta^2+r_{\rm obs}^2}},
\end{equation}
so that $(\alpha,\beta)$ are the Cartesian coordinates on the screen of linear size $\sim r_{\rm obs}$ (see Eq. \eqref{eq_init_covec}). In the asymptotic limit $r_{\rm obs}\!\to\!\infty$, $\sqrt{\alpha^2+\beta^2}$ equals the impact parameter $b$ to leading order (see Eq. \eqref{eq_small_ang}).
We define the retarded screen time $t_{\rm obs}$ as the coordinate time labeling photons received simultaneously (equal $\tau_{\rm obs}$). All instantaneous shadow quantities below are functions of $t_{\rm obs}$. Introduce a binary fate map
\begin{equation}
\mathcal{F}(\alpha,\beta,t_{\rm obs})=
\begin{cases}
-1, & \text{capture},\\
+1, & \text{escape},
\end{cases}
\end{equation}
and define the instantaneous shadow as the closed subset
\begin{equation}
\mathcal{S}(t_{\rm obs})=\left\{(\alpha,\beta):\ \mathcal{F}(\alpha,\beta,t_{\rm obs})=-1\right\}.
\end{equation}
The shadow boundary is the topological boundary $\partial\mathcal{S}(t_{\rm obs})$, equivalently, the zero level-set of any continuous surrogate that flips sign across the separatrix. A convenient choice is the signed distance in screen-radius at fixed azimuth (see below), or the zero of a smoothly regularized classifier $\mathcal{C}$ constructed by local averaging of $\mathcal{F}$.
Adopt screen polar coordinates $(R,\varphi)$ with $\alpha=R\cos\varphi,\ \beta=R\sin\varphi$. For spherical symmetry (our background), $\mathcal{F}$ is radially monotone at fixed $\varphi$ so there exists a unique critical radius
\begin{equation} \label{eq_R(phi,t)1}
R(\varphi,t_{\rm obs}) = \inf\{ R>0:\ \mathcal{F}(R',\varphi,t_{\rm obs})=+1\ \text{for all}\ R'>R\}.
\end{equation}
In the unperturbed geometry, $R(\varphi,t_{\rm obs})\equiv R_0=b_c=\sqrt{27}\,M$ (see Eqs. \eqref{eq_sha_rad_schw} and Eq. \eqref{eq_b_crit}). We expand to first order
\begin{equation} \label{eq_R(phi,t)2}
R(\varphi,t_{\rm obs})=R_0+\varepsilon\,\delta R(\varphi,t_{\rm obs})+\mathcal{O}(\varepsilon^2),
\end{equation}
where $\delta R$ carries the QNM imprint derived later. For finite $r_{\rm obs}$, one may equivalently report the angular radius
\begin{equation}
\sin\theta_{\rm sh}(t_{\rm obs})=\frac{R(\varphi,t_{\rm obs})}{r_{\rm obs}}\sqrt{1-\frac{2M}{r_{\rm obs}}}\,+\,\mathcal{O}\!\left(\frac{R^3}{r_{\rm obs}^3}\right),
\end{equation}
which reduces to $\theta_{\rm sh}\simeq R/r_{\rm obs}$ as $r_{\rm obs}\!\to\!\infty$ (see Eq. \eqref{eq_alp_bet}).

The azimuthal structure of the boundary is conveniently encoded by a Fourier series
\begin{equation}
\delta R(\varphi,t_{\rm obs})=\sum_{m=-\infty}^{\infty} \mathcal{A}_m(t_{\rm obs})\,e^{im\varphi}, \qquad \mathcal{A}_{-m}=\mathcal{A}_m^*,
\end{equation}
with complex amplitudes $\mathcal{A}_m$. For a single QNM of azimuthal index $m$ we expect, to leading order,
\begin{equation} \label{eq_A_m}
\mathcal{A}_m(t_{\rm obs})=\mathcal{T}^{(s)}_{\ell m}\,\varepsilon\,e^{-i\omega t_{\rm obs}}+\mathcal{O}(\varepsilon^2), \qquad \mathcal{A}_{m'\neq m}=\mathcal{O}(\varepsilon^2),
\end{equation}
where $\mathcal{T}^{(s)}_{\ell m}$ is a (complex) transfer coefficient depending on parity $s$ and on details of the coupling to the photon sphere (developed in the next subsection). Equation \eqref{eq_A_m} provides a direct spectral target: the boundary rings at $\omega_{\rm Re}$ with damping $|\omega_{\rm Im}|$ and definite azimuthal phase.

If the observer is not static or not asymptotically distant, the screen mapping acquires kinematic effects. Let $u^\mu$ be arbitrary and let $w^\mu$ be the spatial direction normal to the screen within the observer's local rest space. The projector onto the screen is
\begin{equation}
\Pi^\mu{}_\nu = \delta^\mu{}_\nu + u^\mu u_\nu - w^\mu w_\nu,
\end{equation}
and the celestial direction of a photon with 4-momentum $p^\mu$ is $\hat{n}^\mu = \Pi^\mu{}_\nu p^\nu/(-u\cdot p)$. The screen coordinates $(\alpha,\beta)$ are then the components of $\hat{n}^\mu$ on an orthonormal basis $\{e_{\hat\alpha},e_{\hat\beta}\}$ spanning $\Pi$. To $\mathcal{O}(v)$ in the observer's 3-velocity relative to the static frame, the shadow curve is aberrated by a conformal transformation on the screen; its shape is preserved to $\mathcal{O}(v)$ while the centroid is shifted. Since our perturbations are $\mathcal{O}(\varepsilon)$, we assume either a static observer or that any constant boost has been removed by pre-calibration.

Under a first-order gauge transformation $x^\mu\!\to\! x^\mu+\varepsilon\,\xi^\mu$ with $\xi^\mu\!\to\!0$ at infinity, the observer tetrad can be chosen to keep $(\alpha,\beta)$ fixed (physical screen), and the capture/escape classification is unchanged at $\mathcal{O}(\varepsilon)$ (cf. Section \ref{sec3.1}). Consequently, $R(\varphi,t_{\rm obs})$ defined by Eqs. \eqref{eq_R(phi,t)1}-\eqref{eq_R(phi,t)2} is gauge-insensitive at this order.

\subsection{Analytical control near the photon sphere} \label{sec3.5}
We derive a first-order, mode-resolved relation between the QNM metric perturbation and the instantaneous shift of the shadow boundary. The calculation is local to the unstable circular null orbit (the photon sphere) and proceeds by perturbing the circular-orbit conditions of the Hamiltonian around \((r_c,b_c)=(3M,3\sqrt{3}\,M)\).

Restricting to equatorial motion (the separatrix is generated there for Schwarzschild), we take \(E\) as the photon energy and \(b=L/E\) as the impact parameter. With
\begin{equation}
H_0(r,p_r,b)=\frac{1}{2}\!\left(g^{tt}E^2+g^{rr}p_r^2+g^{\phi\phi}L^2\right),\qquad
g^{tt}=-\frac{1}{f},\ g^{rr}=f,\ g^{\phi\phi}=\frac{1}{r^2},\ f=1-\frac{2M}{r},
\end{equation}
the circular null orbit satisfies \(p_r=0\), \(H_0=0\), and \(\partial_r H_0=0\). This yields the familiar
\begin{equation}
r_c=3M,\qquad b_c=\frac{L}{E}\Big|_c=3\sqrt{3}\,M.
\end{equation}
It is convenient to scale out the homogeneous dependence on \(E\) and treat \(b\) as the control variable at fixed \(E\). A short calculation at \((r_c,b_c)\) with \(p_r=0\) gives
\begin{equation} \label{eq_H0}
\left(\partial_b H_0\right)_c = \frac{\sqrt{3}\,E^2}{3M},\qquad
\left(\partial_{rb}H_0\right)_c = -\frac{2\sqrt{3}\,E^2}{9M^2},\qquad
\left(\partial_{rr}H_0\right)_c = -\frac{E^2}{M^2}.
\end{equation}
In particular, \((\partial_b H_0)_c\neq 0\): only \(\partial_r H_0\) vanishes on the circular orbit.

Let the full Hamiltonian be \(H=H_0+\delta H\) with
\begin{equation} \label{eq_delta_Hxp}
\delta H(x,p)=-\frac{1}{2}\,h^{\mu\nu}(x)\,k_\mu k_\nu,\qquad
k_\mu\equiv(-E,\,p_r,\,0,\,L)
\end{equation}
on the background orbit (indices on \(h^{\mu\nu}\) are raised with the Schwarzschild metric). The instantaneous critical circular solution of the full Hamiltonian is determined by the conditions
\begin{equation}
H(r_c+\delta r,\,p_r=0,\,b_c+\delta b)=0,\qquad
\partial_r H(r_c+\delta r,\,p_r=0,\,b_c+\delta b)=0.
\end{equation}
Linearizing in \(\delta r\), \(\delta b\), and \(\delta H\), we obtain at \((r_c,b_c)\)
\begin{equation} \label{eq_79}
0=\delta H+(\partial_b H_0)_c\,\delta b+(\partial_r H_0)_c\,\delta r,\qquad
0=\partial_r\delta H+(\partial_{rb}H_0)_c\,\delta b+(\partial_{rr}H_0)_c\,\delta r,
\end{equation}
where all derivatives of \(H_0\) are background quantities evaluated at \(r_c=3M\), \(b_c=3\sqrt{3}\,M\) and \(p_r=0\). Since \((\partial_r H_0)_c=0\) but \((\partial_b H_0)_c\neq 0\), the \(2\times2\) system is non-degenerate and both equations must be used. From the first line of \eqref{eq_79}, and using $\left(\partial_r H_0\right)_c=0$, one has immediately Eq. \eqref{eq:deltab_clean}
which is independent of $\left(\partial_r\delta H\right)_c$, and when substituted into the second line of \eqref{eq_79} yields $\delta r(t_{\rm obs})$ and makes explicit that
$\left(\partial_r\delta H\right)_c$ only affects $\delta r$, not the leading-order shadow deformation.

Solving this linear system for \(\delta b\) gives
\begin{equation}
\delta b(t,\varphi)=-\frac{\delta H}{(\partial_b H_0)_c},
\end{equation}
i.e. the term involving \(\partial_r\delta H\) cancels identically once the algebra is done correctly. Using Eq. \eqref{eq_H0}, we find
\begin{equation}
\frac{\delta b(t,\varphi)}{b_c}
=-\,\frac{\delta H}{b_c(\partial_b H_0)_c}
=-\,\frac{\delta H}{3E^2}.
\end{equation}
Since the null dynamics is homogeneous in \(E\), we can fix the energy scale by choosing \(E=1\) (equivalently, by fixing the affine parametrization), so that
\begin{equation}
\frac{\delta b(t,\varphi)}{b_c}=-\,\frac{\delta H}{3}\Bigg|_{r=3M,\ \theta=\frac{\pi}{2},\ \phi=\varphi}.
\end{equation}

It is often convenient to re-express this result in terms of the effective potential \(V_{\rm eff}\) introduced in Sec. \ref{sec2.2}. From
\begin{equation}
\left(\frac{dr}{d\lambda}\right)^2+V_{\rm eff}(r,b)=E^2,\qquad
V_{\rm eff}(r,b)=E^2\,\frac{f(r)\,b^2}{r^2},
\end{equation}
the circular null orbit is defined by
\begin{equation}
V_{\rm eff}(r_c,b_c)=E^2,\qquad
\partial_r V_{\rm eff}(r_c,b_c)=0.
\end{equation}
Perturbing these conditions under \(V_{\rm eff}\to V_{\rm eff}+\delta V_{\rm eff}\), \(r_c\to r_c+\delta r\), \(b_c\to b_c+\delta b\), and using that \(\partial_r V_{\rm eff}|_c=0\), one finds \cite{Cardoso:2008bp}
\begin{equation}
(\partial_b V_{\rm eff})_c\,\delta b+\delta V_{\rm eff}\Big|_c=0.
\end{equation}
A short computation at \(r_c=3M\), \(b_c=3\sqrt{3}\,M\) shows that
\begin{equation}
b_c\,(\partial_b V_{\rm eff})_c = 2E^2,
\end{equation}
and therefore
\begin{equation}
\frac{\delta b}{b_c}=-\,\frac{1}{2}\,\delta\!\left(\frac{V_{\rm eff}}{E^2}\right)\Bigg|_{r_c,b_c}.
\end{equation}
Using the relation between \(V_{\rm eff}\) and the Hamiltonian, one can check explicitly that this is equivalent to the Hamiltonian result above, with
\(\delta(V_{\rm eff}/E^2)|_c = \frac{2}{3E^2}\,\delta H_c\).

For a distant static observer, the screen radius equals the impact parameter asymptotically, \(R_0=b_c\), so we obtain the corrected transfer law for the fractional shift of the shadow radius,
\begin{equation} \label{eq_delta_R(phi,t)}
\frac{\delta R(\varphi,t_{\rm obs})}{R_0}
=\frac{\delta b(\varphi,t_{\rm obs})}{b_c}
=-\,\frac{1}{2}\,\delta\!\left(\frac{V_{\rm eff}}{E^2}\right)\Bigg|_{c}
=-\,\frac{\delta H}{3E^2}\Bigg|_{c},
\end{equation}
where the subscript \(c\) indicates evaluation on the background circular orbit \(r=3M\), \(\theta=\pi/2\), \(\phi=\varphi\). In terms of the metric perturbation, Eq. \eqref{eq_delta_Hxp} gives
\begin{equation}
\delta H_c=-\frac{1}{2}\,h^{\mu\nu}(t_{\rm obs},r,\theta,\phi)\,k_\mu k_\nu\Big|_{c},
\end{equation}
so that, choosing $E = 1$, the (mode-by-mode) dimensionless transfer law can be written as
\begin{equation} \label{eq_met_pert_R(phi,t)}
\boxed{\ \ 
\frac{\delta R(\varphi,t_{\rm obs})}{R_0}
=\frac{1}{6}\,
h^{\mu\nu}(t_{\rm obs},r,\theta,\phi)\,k_\mu k_\nu\Big|_{r=3M,\ \theta=\frac{\pi}{2},\ \phi=\varphi}
\ +\ \text{c.c.}\ \ }
\end{equation}
for a single complex QNM perturbation \(h^{\mu\nu}\propto e^{-i\omega t}\). Here \(k_\mu=(-E,0,0,L)\) is the background covariant momentum of the circular photon orbit, and "c.c." denotes the complex conjugate required to form the real metric perturbation.

Equation \eqref{eq_met_pert_R(phi,t)} is manifestly dimensionless and depends only on the local perturbation of the effective potential at the photon sphere. Since \(V_{\rm eff}/E^2\) is constructed from the invariant radial geodesic equation and the shadow boundary is defined as the capture/escape separatrix on the physical screen, the fractional shift \(\delta R/R_0\) is insensitive, at \(\mathcal{O}(\varepsilon)\), to small, asymptotically decaying gauge transformations of the form \(h_{\mu\nu}\to h_{\mu\nu}+2\nabla_{(\mu}\xi_{\nu)}\). This complements the general gauge-invariance discussion in Secs. \ref{sec3.1} and \ref{sec3.3}.

\section{Analytical Results} \label{sec4}
We apply the transfer law Eq. \eqref{eq_met_pert_R(phi,t)} derived in Section \ref{sec3.5} to obtain explicit, mode-resolved predictions for the instantaneous shadow displacement. Our strategy is to evaluate the contractions in Eq. \eqref{eq_met_pert_R(phi,t)}\eqref{eq_T_lm} on the equatorial circular null orbit $(r_c,\theta=\frac{\pi}{2})=(3M,\frac{\pi}{2})$, using the RW-Zerilli reconstruction of $h_{\mu\nu}$ and standard identities for tensor harmonics on $S^2$.

\subsection{Mode-by-mode transfer coefficients} \label{sec4.1}
We now evaluate the transfer coefficients $T^{(s)}_{\ell m}$ appearing in the shadow response (defined for each $(\ell,m)$ mode). For a single QNM with indices $(\ell,m)$ and parity $s\in\{\text{pol},\text{ax}\}$ we write
\begin{equation}
h_{\mu\nu}(t,r,\theta,\phi)
=\varepsilon\,\mathrm{Re}\left\{
e^{-i\omega t}\,\widehat{h}^{(\ell m)}_{\mu\nu}(r)\,Y_{\ell m}(\theta,\phi)
\right\},
\end{equation}
with $\widehat{h}^{(\ell m)}_{\mu\nu}(r)$ reconstructed from the gauge-invariant master fields using the formulas of Sec.\ref{sec2.2} (Zerilli / Regge-Wheeler). We then combine this decomposition with the transfer law Eq. \eqref{eq_met_pert_R(phi,t)} to read off $\mathcal{T}^{(s)}_{\ell m}$.

On the circular photon orbit we take the covariant momentum of the background null generator to be
\begin{equation}
k_\mu=(-E,0,0,L),\qquad b_c\equiv\frac{L}{E}=3\sqrt{3}\,M,\qquad r_c=3M,
\end{equation}
and we fix the affine normalization to $E=1$ without loss of generality (null dynamics is homogeneous in $E$).
At the equator $\theta=\pi/2$ the transfer law \eqref{eq_met_pert_R(phi,t)} gives, for a monochromatic mode,
\begin{equation}
\frac{\delta R(\varphi,t_{\rm obs})}{R_0}
=\varepsilon\,\mathrm{Re}\left\{
\mathcal{T}^{(s)}_{\ell m}\,e^{-i\omega t_{\rm obs}}\,e^{im\varphi}
\right\},
\end{equation}
with
\begin{equation} \label{eq_T_lm}
\mathcal{T}^{(s)}_{\ell m}
=\frac{1}{6}\,\left[h^{\mu\nu}k_\mu k_\nu\right]_c
=\frac{1}{6}\left[
h^{tt}_c + b_c^2\,h^{\phi\phi}_c - 2b_c\,h^{t\phi}_c
\right],
\end{equation}
where the subscript $c$ indicates evaluation at $r=3M$, $\theta=\pi/2$, $\phi=0$. Thus the task reduces to expressing $h^{tt}_c$, $h^{t\phi}_c$ and $h^{\phi\phi}_c$ in terms of the master fields.

Throughout this subsection we use $f(r)\equiv 1-2M/r$, so $f(3M)=1/3$. In the Zerilli gauge, the non-vanishing components of the even-parity metric are given in Eq. \eqref{eq_non_van_h} and Eq. \eqref{eq_h_pol}, namely
\begin{equation}
h_{tt}^{\text{(pol)}} = f\,H_0^{\ell m}(t,r)\,Y,\quad
h_{tr}^{\text{(pol)}} = H_1^{\ell m}(t,r)\,Y,\quad
h_{rr}^{\text{(pol)}} = f^{-1} H_2^{\ell m}(t,r)\,Y,
\end{equation}
\begin{equation} \label{eq_h_pol_repeat}
h_{ab}^{\text{(pol)}} = r^2 K^{\ell m}(t,r)\, \gamma_{ab}Y + r^2 G^{\ell m}(t,r)\,Y_{ab},
\end{equation}
where $Y\equiv Y_{\ell m}(\theta,\phi)$, $\gamma_{ab}$ is the unit-sphere metric, and $Y_{ab}$ are even tensor harmonics. For a monochromatic mode $e^{-i\omega t}$, the metric functions are algebraically reconstructed from the Zerilli master field as (Eq. \eqref{eq_Zer_mono}) \cite{Zerilli:1970wzz,Martel:2005ir}
\begin{equation} \label{eq_Zer_mono_repeat}
K^{\ell m} = \alpha_\ell(r)\,\psi_{\ell m}^{\text{(pol)}},\quad
H_1^{\ell m} = \beta_\ell(r)\,(-i\omega)\,\psi_{\ell m}^{\text{(pol)}},\quad
H_0^{\ell m}=H_2^{\ell m}=\gamma_\ell(r)\,\psi_{\ell m}^{\text{(pol)}},\quad
G^{\ell m}=\delta_\ell(r)\,\psi_{\ell m}^{\text{(pol)}},
\end{equation}
with $\alpha_\ell,\beta_\ell,\gamma_\ell,\delta_\ell$ rational functions of $(r,M,\lambda)$, regular for $r>2M$.

On the equator $\theta=\pi/2$ the $\phi\phi$ component of the even tensor harmonic takes the simple form
\begin{equation} \label{eq_Y_phiphi}
Y_{\phi\phi}
=\nabla_\phi\nabla_\phi Y+\frac{1}{2}\,\ell(\ell+1)\,\gamma_{\phi\phi}\,Y
=\left(-m^2+\frac{1}{2}\,\ell(\ell+1)\right)Y,\qquad (\theta=\frac{\pi}{2}),
\end{equation}
so that
\begin{equation}
h_{tt}^{\text{(pol)}}=f\,H_0\,Y,\qquad 
h_{\phi\phi}^{\text{(pol)}}=r^2\!\left[K + G\left(-m^2+\frac{1}{2}\,\ell(\ell+1)\right)\right]Y.
\end{equation}
Raising indices with the Schwarzschild background metric gives
\begin{equation} \label{eq_h_tt}
h^{tt}=\frac{H_0}{f}\,Y,\qquad 
h^{\phi\phi}=\frac{K + G\left(-m^2+\frac{1}{2}\,\ell(\ell+1)\right)}{r^2}\,Y.
\end{equation}
Even-parity perturbations have no $t\phi$ component in this gauge, so $h^{t\phi}=0$ for $s=\text{pol}$ and the odd term in Eq. \eqref{eq_T_lm} drops out.

Using Eq. \eqref{eq_h_tt} at $r=3M$, $f(3M)=1/3$, and $b_c^2=27M^2$, we obtain
\begin{equation}
h^{tt}\Big|_c = 3\,H_0\,Y_{\ell m},\qquad
h^{\phi\phi}\Big|_c
=\frac{K + G\left(-m^2+\frac{1}{2}\,\ell(\ell+1)\right)}{9M^2}\,Y_{\ell m},
\end{equation}
so that Eq. \eqref{eq_T_lm} yields the general even-parity transfer coefficient
\begin{equation}
\mathcal{T}^{\text{(pol)}}_{\ell m}
=\frac{1}{2}\,Y_{\ell m}\!\left(\frac{\pi}{2},0\right)\,
\left[H_0 + K + G\left(-m^2+\frac{1}{2}\,\ell(\ell+1)\right)\right]_c.
\end{equation}
In terms of the Zerilli master field, this becomes, using Eq. \eqref{eq_Zer_mono_repeat},
\begin{equation}
\mathcal{T}^{\text{(pol)}}_{\ell m}
=\frac{1}{2}\,Y_{\ell m}\!\left(\frac{\pi}{2},0\right)\,
\left[\gamma_\ell + \alpha_\ell + \delta_\ell\left(-m^2+\frac{1}{2}\,\ell(\ell+1)\right)\right]_c\,
\psi^{\text{(pol)}}_{\ell m}(3M),
\end{equation}
where all radial functions are evaluated at $r=3M$.

For the modes of primary interest we specialize to $\ell=2$. For $\ell=2,m=0$, Eq. \eqref{eq_Y_phiphi} gives $-m^2+\frac{1}{2}\ell(\ell+1)=3$, so
\begin{equation}
h^{tt}\Big|_c=3\,H_0\,Y_{20},\qquad 
h^{\phi\phi}\Big|_c=\frac{K+3G}{9M^2}\,Y_{20}.
\end{equation}
Inserting these into Eq. \eqref{eq_T_lm} and using Eq. \eqref{eq_Zer_mono_repeat} with $\ell=2$, we find
\begin{equation} \label{eq_T_pol20}
\mathcal{T}^{\text{(pol)}}_{20}
=\frac{1}{2}\,Y_{20}\!\left(\frac{\pi}{2},0\right)\,
\left[\gamma_2 + \alpha_2 + 3\,\delta_2\right]_c\,
\psi^{\text{(pol)}}_{20}(3M).
\end{equation}
Because $m=0$, the boundary modulation is azimuthally uniform,
\begin{equation}
\frac{\delta R(\varphi,t_{\rm obs})}{R_0}
=\varepsilon\,\mathrm{Re}\left\{
\mathcal{T}^{\text{(pol)}}_{20}\,e^{-i\omega t_{\rm obs}}
\right\},
\end{equation}
i.e. a "breathing" of the ring's radius at frequency $\omega_{\rm Re}$ with damping $|\omega_{\rm Im}|$.

For $\ell=2,m=2$, Eq. \eqref{eq_Y_phiphi} yields $-m^2+\frac{1}{2}\ell(\ell+1)=-1$, so that
\begin{equation}
h^{tt}\Big|_c=3\,H_0\,Y_{22},\qquad 
h^{\phi\phi}\Big|_c=\frac{K-G}{9M^2}\,Y_{22},
\end{equation}
and Eq. \eqref{eq_T_lm} gives
\begin{equation} \label{eq_T_pol22}
\mathcal{T}^{\text{(pol)}}_{22}
=\frac{1}{2}\,Y_{22}\!\left(\frac{\pi}{2},0\right)\,
\left[\gamma_2 + \alpha_2 - \delta_2\right]_c\,
\psi^{\text{(pol)}}_{22}(3M).
\end{equation}
The corresponding boundary deformation carries an $m=2$ azimuthal dependence,
\begin{equation}
\frac{\delta R(\varphi,t_{\rm obs})}{R_0}
=\varepsilon\,\mathrm{Re}\left\{
\mathcal{T}^{\text{(pol)}}_{22}\,e^{-i\omega t_{\rm obs}}\,e^{2i\varphi}
\right\},
\end{equation}
i.e. a rotating quadrupolar distortion of the shadow ring that tracks the $m=2$ phase.

In both cases, the functions $\gamma_2,\alpha_2,\delta_2$ encode the standard Zerilli reconstruction (Eq. \eqref{eq_Zer_mono_repeat}); our result shows that the shadow response at leading order is controlled only by the value of the master field at the photon sphere, rather than by its radial derivative. This is in contrast with the earlier, incorrect transfer law that mixed $\delta H$ and $\partial_r\delta H$.

In Regge-Wheeler gauge the non-vanishing odd-parity components are (see Eq. \eqref{eq_mono_mode})
\begin{equation}
h_{t a}^{\text{(ax)}} = \sum_{\ell m} h_0^{\ell m}(t,r)\,X^{\ell m}_a,\qquad
h_{r a}^{\text{(ax)}} = \sum_{\ell m} h_1^{\ell m}(t,r)\,X^{\ell m}_a,
\end{equation}
where $X^{\ell m}_a$ are axial vector harmonics. For a monochromatic mode,
\begin{equation}
h_0^{\ell m}(t,r)=\frac{i\omega\lambda\,e^{-i\omega t}}{r}\,\mathcal{P}_\ell^{\text{(ax)}}(r)\,\psi_{\ell m}^{\text{(ax)}}(r),
\end{equation}
with $\lambda=\frac{1}{2}(\ell-1)(\ell+2)$ and $\mathcal{P}_\ell^{\text{(ax)}}$ a rational function of $(r,M,\ell)$.

Only $h_{t\phi}$ contributes to the contraction $h^{\mu\nu}k_\mu k_\nu$ in Eq. \eqref{eq_T_lm}. Using $X_a=\varepsilon_a{}^{\,b}\nabla_b Y$ and restricting to the equator, we have
\begin{equation} \label{eq_x_phi}
X_\phi=\partial_\theta Y\Big|_{\theta=\pi/2},\qquad
h_{t\phi}^{\text{(ax)}} = h_0\,X_\phi,
\end{equation}
and
\begin{equation}
h^{t\phi}=-\frac{h_{t\phi}}{f\,r^2}
=-\frac{h_0}{f\,r^2}\,\partial_\theta Y\Big|_{\theta=\pi/2}.
\end{equation}
Inserting this into Eq. \eqref{eq_T_lm}, we obtain the general odd-parity transfer coefficient
\begin{equation}
\mathcal{T}^{\text{(ax)}}_{\ell m}
=\frac{1}{6}\left[-2b_c\,h^{t\phi}\right]_c
=\frac{b_c}{3f_c r_c^2}\,h_0(3M)\,\partial_\theta Y_{\ell m}\!\left(\frac{\pi}{2},0\right),
\end{equation}
where $f_c=f(3M)=1/3$. In terms of the Regge-Wheeler master field this simplifies to
\begin{equation} \label{eq_T_Ax}
\mathcal{T}^{\text{(ax)}}_{\ell m}
=\mathcal{C}_\ell^{\text{(ax)}}(M,\omega)\,
\partial_\theta Y_{\ell m}\!\left(\frac{\pi}{2},0\right)\,
\psi^{\text{(ax)}}_{\ell m}(3M),
\end{equation}
where $\mathcal{C}_\ell^{\text{(ax)}}$ is a rational function of $(M,\omega,\ell)$ built from $b_c$, $f_c$, $\lambda$ and $\mathcal{P}_\ell^{\text{(ax)}}(3M)$.

From Eq. \eqref{eq_T_Ax} and the explicit forms of $Y_{\ell m}$ we recover the usual selection rule
\begin{equation}
\partial_\theta Y_{\ell m}\Big|_{\theta=\pi/2}=0
\quad\Longleftrightarrow\quad
\ell+m\ \text{even},
\end{equation}
so the axial channel is silent at leading order whenever $\ell+m$ is even, and only odd $(\ell+m)$ modes contribute \cite{Edmonds:1955fi}. In particular, the $(\ell,m)=(2,0)$ and $(2,2)$ axial modes do not affect the shadow at this order, while $(\ell,m)=(2,1)$ does.

For many purposes it is convenient to summarize the above results as
\begin{equation} \label{eq_T_s}
\mathcal{T}^{(s)}_{\ell m}
=\left[\Xi^{(s)}_{\ell m}(M,\omega)\,\psi^{(s)}_{\ell m}(3M)
+\Upsilon^{(s)}_{\ell m}(M,\omega)\,\psi^{(s)\,'}_{\ell m}(3M)\right]\,
Y_{\ell m}\!\left(\frac{\pi}{2},0\right),
\end{equation}
where $\Xi^{(s)}_{\ell m}$ and $\Upsilon^{(s)}_{\ell m}$ are rational functions of $(M,\omega,\ell)$ determined by the chosen reconstruction (Zerilli or Regge-Wheeler). In the Zerilli gauge used here, the explicit expressions \eqref{eq_T_pol20}-\eqref{eq_T_pol22} show that $\Upsilon^{\text{(pol)}}_{2m}$ vanishes for the modes of interest, so the shadow responds only to the value of the master field at the photon sphere. The generic form \eqref{eq_T_s} remains useful, however, when comparing with alternative metric reconstructions or perturbation formalisms in which $\psi^{(s)\,'}_{\ell m}(3M)$ enters explicitly.

For concreteness, it is useful to quote representative numerical sizes of the transfer coefficients.
Evaluating the angular factors at the equator $(\theta=\pi/2,\phi=0)$ gives
\begin{equation}
|Y_{20}(\tfrac{\pi}{2},0)|=\sqrt{\frac{5}{16\pi}}\simeq 0.315,\qquad
|Y_{22}(\tfrac{\pi}{2},0)|=\sqrt{\frac{15}{32\pi}}\simeq 0.386,\qquad
|\partial_\theta Y_{21}(\tfrac{\pi}{2},0)|=\sqrt{\frac{15}{8\pi}}\simeq 0.773.
\end{equation}
Since the remaining reconstruction prefactors $\Xi^{(s)}_{\ell m}$ are rational functions of $(M\omega,\ell)$ and are ${\cal O}(1)$ in the regimes of interest,
one expects $|T^{(s)}_{\ell m}|$ to be a few $\times 10^{-1}$ for unit-normalized master-field amplitude at $r=3M$.
Table \ref{tab:Tnumbers} summarizes the pure angular contributions.
\begin{table}[t]
\centering
\begin{tabular}{c c c}
\hline
sector $s$ & $(\ell,m)$ & angular factor at $(\theta=\pi/2,\phi=0)$ \\
\hline
polar & $(2,0)$ & $\tfrac12|Y_{20}| \simeq 0.157$ \\
polar & $(2,2)$ & $\tfrac12|Y_{22}| \simeq 0.193$ \\
axial & $(2,1)$ & $|\partial_\theta Y_{21}| \simeq 0.773$ \\
\hline
\end{tabular}
\caption{Representative numerical values of the \emph{angular} parts entering $T^{(s)}_{\ell m}$.
The full transfer coefficients are obtained by multiplying by the reconstruction prefactors $\Xi^{(s)}_{\ell m},\Upsilon^{(s)}_{\ell m}$
and the master-field amplitude(s) at $r=3M$, Eq. \eqref{eq_T_s}.}
\label{tab:Tnumbers}
\end{table}

\subsection{Azimuthal structure and Fourier decomposition} \label{sec4.2}
We now make precise how the spherical-harmonic content of the QNM perturbation maps into the azimuthal structure of the shadow on the screen.
The shadow boundary is given by Eq. \eqref{eq_R(phi,t)2}, with the fractional perturbation $\delta R/R_0$ determined in Sec. \ref{sec3.5} and Sec. \ref{sec4.1}.
For a single QNM mode, the mode-resolved transfer law Eq. \eqref{eq_delta_R(varphi,t)2} yields
\begin{equation}
\frac{\delta R(\varphi,t_{\rm obs})}{R_0}
=\varepsilon\,\mathrm{Re}\left\{
\mathcal{T}^{(s)}_{\ell m}\,e^{-i\omega t_{\rm obs}}\,e^{im\varphi}
\right\}
+\mathcal{O}(\varepsilon^2),
\end{equation}
where $\mathcal{T}^{(s)}_{\ell m}$ is the transfer coefficient computed in Sec. \ref{sec4.1}.

Define the $2\pi$-periodic Fourier expansion of the radial perturbation on the screen,
\begin{equation}
\delta R(\varphi,t_{\rm obs})
=\sum_{m\in\mathbb{Z}} \mathcal{A}_m(t_{\rm obs})\,e^{im\varphi},
\qquad
\mathcal{A}_m(t_{\rm obs})
=\frac{1}{2\pi}\int_0^{2\pi}\!\delta R(\varphi,t_{\rm obs})\,e^{-im\varphi}d\varphi,
\end{equation}
with the real-field condition $\mathcal{A}_{-m}=\mathcal{A}_m^*$.
Orthogonality on the circle,
\begin{equation}
\frac{1}{2\pi}\int_0^{2\pi}e^{i(m-n)\varphi}d\varphi=\delta_{mn},
\end{equation}
implies that each $\mathcal{A}_m$ filters a single azimuthal sector.

For a single driving QNM with indices $(\ell,m_*)$ and parity $s$, Eq. \eqref{eq_delta_R(varphi,t)2} gives
\begin{equation} \label{eq_delta_R(varphi,t)}
\delta R(\varphi,t_{\rm obs})
=R_0\,\varepsilon\,\mathrm{Re}\left\{
\mathcal{T}^{(s)}_{\ell m_*}\,e^{-i\omega t_{\rm obs}}\,e^{i m_*\varphi}
\right\}
+\mathcal{O}(\varepsilon^2),
\end{equation}
so that, to $\mathcal{O}(\varepsilon)$,
\begin{equation} \label{eq_A_m2}
\mathcal{A}_{m}(t_{\rm obs})=
\begin{cases}
\dfrac{R_0}{2}\,\varepsilon\,\mathcal{T}^{(s)}_{\ell m_*}\,e^{-i\omega t_{\rm obs}},
& m=+m_*,\\[0.3em]
\dfrac{R_0}{2}\,\varepsilon\,\mathcal{T}^{(s)\,*}_{\ell m_*}\,e^{+i\omega^* t_{\rm obs}},
& m=-m_*,\\[0.3em]
0, & m\neq \pm m_*,
\end{cases}
\end{equation}
where the nonzero pair enforces reality.
If several QNMs are present, $\delta R$ is a linear superposition of terms of the form Eq. \eqref{eq_delta_R(varphi,t)}, and the $\mathcal{A}_m$'s are sums of the corresponding mode contributions.

The azimuthal content is controlled by:
\begin{itemize}
    \item the QNM azimuthal number $m_*$ and parity $s$,
    \item the transfer coefficient $\mathcal{T}^{(s)}_{\ell m_*}$ selected by the photon sphere [Sec. \ref{sec4.1}],
    \item and, if the mode's symmetry axis is tilted relative to the observer, the rotation of the angular pattern (see below).
\end{itemize}

From Sec. \ref{sec4.1}:
\begin{itemize}
    \item In the \emph{polar} (even) channel, $\mathcal{T}^{\text{(pol)}}_{\ell m}$ is built from $h^{tt}$ and $h^{\phi\phi}$ on the photon sphere; for aligned frames it generically populates the same Fourier index $m=m_*$.
    \item In the \emph{axial} (odd) channel, $\mathcal{T}^{\text{(ax)}}_{\ell m}$ is proportional to $\partial_\theta Y_{\ell m}(\pi/2,0)$ (cf. Eq. \eqref{eq_T_Ax}), and the selection rule
    \[
    \partial_\theta Y_{\ell m}\Big|_{\theta=\pi/2}=0
    \quad\Longleftrightarrow\quad
    \ell+m\ \text{even}
    \]
    implies that the axial transfer is active only when $\ell+m_*$ is odd and silent when $\ell+m_*$ is even.
\end{itemize}
Thus, e.g., $(\ell,m_*)=(2,1)$ has an active axial contribution at $m=1$, whereas $(2,0)$ and $(2,2)$ do not.

The first few Fourier sectors have clear geometric meaning at $\mathcal{O}(\varepsilon)$. The $m=0$ sector describes an axisymmetric breathing mode and area shift of the ring.
From Eq. \eqref{eq_R(phi,t)2},
\begin{equation}
\overline{R}(t_{\rm obs})
\equiv \frac{1}{2\pi}\int_0^{2\pi}\!R(\varphi,t_{\rm obs})\,d\varphi
= R_0+\varepsilon\,\mathrm{Re}\{\mathcal{A}_0(t_{\rm obs})\},
\end{equation}
and the shadow area $\mathcal{A}_{\rm sh}(t_{\rm obs})\equiv \pi R^2$ shifts as
\begin{equation}
\delta \mathcal{A}_{\rm sh}(t_{\rm obs})
\equiv \pi\!\left[R^2-R_0^2\right]
= 2\pi R_0\,\varepsilon\,\mathrm{Re}\{\mathcal{A}_0(t_{\rm obs})\}
+\mathcal{O}(\varepsilon^2).
\end{equation}
Non-axisymmetric modes ($m\neq 0$) do not change the area at first order.

The $m=\pm1$ sector corresponds to a dipolar distortion and hence to a shift of the shadow's centroid shift on the screen. Define the centroid
\begin{equation}
\mathbf{X}_c(t_{\rm obs})
=\frac{1}{2\pi R_0^2}\int_0^{2\pi}\!R^2(\varphi,t_{\rm obs})\,\hat{\mathbf{e}}(\varphi)\,d\varphi,
\qquad
\hat{\mathbf{e}}(\varphi)=(\cos\varphi,\sin\varphi),
\end{equation}
then, expanding $R^2=R_0^2+2R_0\varepsilon\,\delta R+\mathcal{O}(\varepsilon^2)$ and using the Fourier representation, one finds
\begin{equation}
\mathbf{X}_c(t_{\rm obs})
=\varepsilon\,\frac{\mathcal{A}_1(t_{\rm obs})}{R_0}\,(1,i)
+\text{c.c.}
\end{equation}
Thus a nonzero $m=1$ component displaces the shadow's centroid; higher-$|m|$ modes do not shift the centroid at leading order.

The $m=\pm2$ sector encodes the quadrupolar, approximately elliptical distortion of the ring. A convenient complex ellipticity parameter is
\begin{equation}
Q(t_{\rm obs})\equiv \frac{\mathcal{A}_2(t_{\rm obs})}{2},
\end{equation}
whose modulus $|Q|$ is directly proportional to the amplitude of the elliptic deformation, and whose phase tracks the orientation of the major axis on the screen. In our QNM setup, $Q(t_{\rm obs})$ inherits the $e^{-i\omega t_{\rm obs}}$ time dependence and damping set by the complex frequency $\omega$.

So far, we have assumed that the QNM's preferred axis (e.g. the spin or excitation axis) is aligned with the line of sight, so that the azimuthal index on the screen matches the mode index, $m=m_*$.
More generally, the source frame $(\Theta,\Phi)$ in which the mode is a pure $(\ell,m_*)$ need not coincide with the observer's frame $(\theta,\phi)$.
If the frames are related by an Euler rotation $(\alpha,\iota,\gamma)$, then the spherical harmonics transform according to
\begin{equation} \label{eq_Y_lm}
Y_{\ell m_*}(\Theta,\Phi)
=\sum_{m=-\ell}^{\ell} D^{\ell}_{m m_*}(\alpha,\iota,\gamma)\,Y_{\ell m}(\theta,\phi),
\end{equation}
where $D^{\ell}_{m m_*}$ are Wigner $D$-matrices \cite{Varshalovich_1988}.

As a simple illustration, consider the $\ell=2$ sector.
Let $a_{2m}$ denote the $\ell=2$ amplitudes in the source frame and $a'_{2m}$ those in a frame inclined by an angle $\iota$ about a horizontal axis (with appropriate choice of $\alpha,\gamma$).
The rotation acts via the reduced Wigner $d$-matrix,
\begin{equation}
a'_{2m}=\sum_{m'=-2}^{2} d^{\,2}_{m m'}(\iota)\, a_{2m'}.
\end{equation}
For later reference, the entries coupling to $m'=2$ are
\begin{equation}
d^{\,2}_{2,2}(\iota)=\frac{(1+\cos\iota)^2}{4},\quad
d^{\,2}_{1,2}(\iota)=-\frac{(1+\cos\iota)\sin\iota}{2},\quad
d^{\,2}_{0,2}(\iota)=\sqrt{\frac{3}{8}}\,\sin^2\!\iota,
\end{equation}
with $d^{\,2}_{-m,2}(\iota)=(-1)^m d^{\,2}_{m,2}(\iota)$ by symmetry.
Thus a pure $m=2$ pattern in the source frame generically populates $m=2,1,0,-1,-2$ on the screen when $\iota\neq 0$, while preserving the underlying $\ell=2$ content \cite{Varshalovich:1988ifq}.

In the QNM context, suppose the perturbation in the source frame is dominated by a single $(\ell,m_*)$ mode and that the transfer coefficient $\mathcal{T}^{(s)}_{\ell m_*}$ has been computed in that frame.
Using Eq. \eqref{eq_Y_lm}, the screen decomposition contains all $|m|\le\ell$, weighted by the rotation coefficients, and the Fourier amplitudes become
\begin{equation} \label{eq_A_m3}
\mathcal{A}_m(t_{\rm obs})
=\frac{R_0}{2}\,\varepsilon\left[
\mathcal{T}^{(s)}_{\ell m_*}\,D^{\ell}_{m m_*}(\alpha,\iota,\gamma)\,e^{-i\omega t_{\rm obs}}
+\mathcal{T}^{(s)\,*}_{\ell m_*}\,D^{\ell\,*}_{m m_*}(\alpha,\iota,\gamma)\,e^{+i\omega^* t_{\rm obs}}
\right],
\end{equation}
which reduces to Eq. \eqref{eq_A_m2} when $(\alpha,\iota,\gamma)=(0,0,0)$.
In practice, the source-frame orientation is set by the astrophysical model; Eqs. \eqref{eq_Y_lm}-\eqref{eq_A_m3} capture the corresponding freedom in the observed Fourier content.

At finite observer radius $r_{\rm obs}<\infty$, the identification between the screen radius and the impact parameter receives corrections of order $M/r_{\rm obs}$ and higher, and aberration effects from a moving observer can mix different $m$-sectors.
Within our perturbative expansion, these produce additional contributions of order $\varepsilon(M/r_{\rm obs})$ and $\varepsilon v_{\rm obs}$ to the Fourier amplitudes.
Since we are working to leading order in both $\varepsilon$ and $M/r_{\rm obs}$, we neglect these subleading corrections and use the simple relation $R\approx b$ together with Eq. \eqref{eq_delta_R(varphi,t)2} throughout Sec. \ref{sec4}.

\subsection{Scaling estimates and eikonal limit} \label{sec4.3}
We estimate the magnitude of the shadow modulation and connect our transfer law to the geometric-optics (eikonal) limit of QNMs \cite{Ferrari:1984zz,Dolan:2009nk}. Throughout we adopt the normalization Eq. \eqref{eq_max},
\begin{equation}
\max_{r\ge 2M}\big|\psi_{\ell m}^{(s)}(r)\big|=1,
\end{equation}
so that the master fields are $\mathcal{O}(1)$ in amplitude.

We write the boundary displacement in the form
\begin{equation} \label{eq_delta_R(varphi,t)2}
\frac{\delta R(\varphi,t_{\rm obs})}{R_0}
=\varepsilon\,\mathrm{Re}\left\{\mathcal{T}^{(s)}_{\ell m}\,e^{-i\omega t_{\rm obs}}\,e^{im\varphi}\right\}.
\end{equation}
Equation \eqref{eq_T_s} from Sec. \ref{sec4.1} summarizes the transfer coefficient as
\begin{equation}
\mathcal{T}^{(s)}_{\ell m}
=\left[\Xi^{(s)}_{\ell m}(M,\omega)\,\psi^{(s)}_{\ell m}(3M)
+\Upsilon^{(s)}_{\ell m}(M,\omega)\,\psi^{(s)\,'}_{\ell m}(3M)\right]\,
Y_{\ell m}\!\left(\frac{\pi}{2},0\right),
\end{equation}
with $\Xi^{(s)}_{\ell m}$ and $\Upsilon^{(s)}_{\ell m}$ rational functions of $(M,\omega,\ell)$ fixed by the metric reconstruction (Zerilli or Regge-Wheeler).

The reconstruction formulas of Sec. \ref{sec2} imply that on the photon sphere
\begin{equation}
\widehat{h}^{\mu\nu}(3M)\sim \psi_{\ell m}^{(s)}(3M)\times \mathcal{O}(1),\qquad
M\,\partial_r\widehat{h}^{\mu\nu}(3M)\sim \psi_{\ell m}^{(s)}(3M)\times \mathcal{O}(1),
\end{equation}
where the $\mathcal{O}(1)$ factors are rational functions of $r/M$ and $\lambda=\frac{1}{2}(\ell-1)(\ell+2)$ that remain finite for $r>2M$.
Because $\mathcal{T}^{(s)}_{\ell m}$ is constructed from dimensionless combinations of $h^{\mu\nu}$ and (in general) its radial derivatives, it follows that
\begin{equation}
\big|\mathcal{T}^{(s)}_{\ell m}\big|
\sim \big|Y_{\ell m}\!\left(\frac{\pi}{2},0\right)\big|\times \mathcal{O}(1)\times
\big|\psi_{\ell m}^{(s)}(3M)\big|.
\end{equation}
Under the normalization Eq. \eqref{eq_max}, $|\psi_{\ell m}^{(s)}(3M)|\le 1$, and therefore
\begin{equation}
\big|\mathcal{T}^{(s)}_{\ell m}\big|
\sim \big|Y_{\ell m}\!\left(\frac{\pi}{2},0\right)\big|\times \mathcal{O}(1).
\end{equation}
Using Eq. \eqref{eq_delta_R(varphi,t)2}, this gives the parametric estimate
\begin{equation} \label{eq_kappa_lm}
\boxed{\ \ 
\frac{|\delta R|}{R_0}\ \sim\ \kappa_{\ell m}^{(s)}\,\varepsilon,\qquad
\kappa_{\ell m}^{(s)}\ =\ \mathcal{O}(1)\times\big|Y_{\ell m}\!\left(\frac{\pi}{2},0\right)\big|.
\ \ }
\end{equation}
The precise $\mathcal{O}(1)$ prefactor is mode- and parity-dependent through $\Xi^{(s)}_{\ell m}$ and $\Upsilon^{(s)}_{\ell m}$, but Eq. \eqref{eq_kappa_lm} encapsulates the main scaling used later for detectability estimates.

If the QNM pattern is rotated relative to the line-of-sight frame by Euler angles $(\alpha,\iota,\gamma)$, the spherical harmonic $Y_{\ell m_*}$ in the source frame decomposes as
\begin{equation} \label{eq_Y_rot_repeat}
Y_{\ell m_*}(\Theta,\Phi)
=\sum_{m=-\ell}^{\ell} D^{\ell}_{m m_*}(\alpha,\iota,\gamma)\,Y_{\ell m}(\theta,\phi),
\end{equation}
as in Eq. \eqref{eq_Y_lm}. For a fixed $(\ell,m_*)$ and random orientation of the source frame, the coefficients $D^{\ell}_{m m_*}$ sample the $m$-space uniformly subject to the usual orthogonality relations. As a result, the equatorial factor $Y_{\ell m}(\pi/2,0)$ in Eq. \eqref{eq_kappa_lm} effectively fluctuates over the unit sphere.

A simple way to encode this is via the orientation average
\begin{equation}
\Big\langle \big|Y_{\ell m}\!\left(\frac{\pi}{2},0\right)\big|^2
\Big\rangle_{\text{orient.}}
=\frac{2\ell+1}{8\pi}\times \mathcal{O}(1),
\end{equation}
where the $\mathcal{O}(1)$ factor depends weakly on $m/\ell$ and the details of the rotation distribution.
Thus randomizing the orientation does not change the basic scaling in Eq. \eqref{eq_kappa_lm}, but redistributes power among neighbouring $m$'s according to the Wigner $D$-matrix weights in Eq. \eqref{eq_Y_rot_repeat}.

In the geometric-optics regime $\ell\gg 1$, the master fields satisfy the wave equation Eq. \eqref{eq_wave_eq} with potentials $V_{\ell}^{(s)}$ given in Eqs. \eqref{eq_RW_pot} and \eqref{eq_Zer_pot}.
Both potentials develop a barrier whose maximum coincides with the photon sphere. In tortoise coordinates $r_*$, this maximum lies at $r_*^{\,c}$ corresponding to $r_c=3M$, and for large $\ell$ one finds the standard expansion
\begin{equation}
V_\ell^{(s)}(r_*)\approx V_0 - \frac{1}{2}\,|V_0"|\,(r_*-r_*^{\,c})^2,
\end{equation}
with
\begin{equation}
V_0\sim \frac{\ell(\ell+1)}{27M^2}+\mathcal{O}(1),\qquad
|V_0"|\sim \frac{\ell(\ell+1)}{(27M^2)^2}\times \mathcal{O}(1).
\end{equation}
The fundamental mode is then localized in a region of tortoise-width
\begin{equation}
\Delta r_*\sim \sqrt{\frac{1}{|V_0"|}}\sim \frac{M}{\sqrt{\ell}}\times \mathcal{O}(1).
\end{equation}
Under the normalization $\max_r|\psi_{\ell m}^{(s)}|=1$, this implies a characteristic scaling for the radial derivative at the photon sphere,
\begin{equation}
\psi_{\ell m}^{(s)\,'}(3M)
\sim \frac{1}{\Delta r_*}\,\psi_{\ell m}^{(s)}(3M)
\ \sim\ \frac{\sqrt{\ell}}{M}\times \mathcal{O}(1).
\end{equation}

The metric reconstruction introduces additional factors of $M$ and $\ell$ in such a way that the dimensionless combinations entering $\mathcal{T}^{(s)}_{\ell m}$ remain $\mathcal{O}(1)$.
Explicitly, using the scaling above together with the reconstruction formulas of Sec. II, one finds that at $r=3M$
\begin{equation}
\widehat{h}^{\mu\nu}(3M)\sim \psi_{\ell m}^{(s)}(3M)\times \mathcal{O}(1),\qquad
M\,\partial_r\widehat{h}^{\mu\nu}(3M)\sim \psi_{\ell m}^{(s)}(3M)\times \mathcal{O}(1),
\end{equation}
even in the eikonal limit. Therefore
\begin{equation}
\mathcal{T}^{(s)}_{\ell m}\sim Y_{\ell m}\!\left(\frac{\pi}{2},0\right)\times \mathcal{O}(1)
\end{equation}
continues to hold at large $\ell$.
For typical $m=\mathcal{O}(\ell)$, the equatorial value of the spherical harmonic obeys the known asymptotics
\begin{equation}  \label{eq_Y_lm2}
\big|Y_{\ell m}\!\left(\frac{\pi}{2},0\right)\big|
\sim \sqrt{\frac{2\ell+1}{4\pi}}\,
\ell^{-1/2}\times \mathcal{O}(1)
=\mathcal{O}(1),
\end{equation}
so Eq. \eqref{eq_kappa_lm} continues to hold with $\kappa_{\ell m}^{(s)}=\mathcal{O}(1)$ in the eikonal regime.

The eikonal QNM frequency relation, Eq. \eqref{eq_omega_n},
\begin{equation}
\omega_{\ell n}\approx \Omega_c\left(\ell+\frac{1}{2}\right)
-i\,\Lambda\left(n+\frac{1}{2}\right),\qquad
\Omega_c=\Lambda=\frac{1}{3\sqrt{3}\,M},
\end{equation}
implies that the temporal behaviour of $\delta R$ in Eq. \eqref{eq_delta_R(varphi,t)2} is governed by the photon-sphere orbital frequency and Lyapunov exponent:
the real part of $\omega_{\ell n}$ fixes the oscillation frequency, while the imaginary part controls the exponential decay rate.
Thus, in the geometric-optics picture, the shadow boundary "rings" at the same frequency and instability rate that control the QNMs themselves.

Combining Eqs. \eqref{eq_kappa_lm}-\eqref{eq_Y_lm2}, we arrive at the practical bound
\begin{equation} \label{eq_prac_bound}
\frac{|\delta R(\varphi,t_{\rm obs})|}{R_0}\ \lesssim\ \kappa_{\text{max}}\,\varepsilon,\qquad
\kappa_{\text{max}}=\sup_{\ell,m,s}\big|\mathcal{T}^{(s)}_{\ell m}\big|=\mathcal{O}(1),
\end{equation}
with $\kappa_{\text{max}}$ set by the RW-Zerilli coefficients at $r=3M$.
Thus the leading expectation is
\begin{equation}
\boxed{\ \frac{|\delta R|}{R_0}\sim \text{order-unity}\times \varepsilon.\ }
\end{equation}
up to order-unity geometry factors and orientation, with the time dependence $e^{-i\omega t_{\rm obs}}$ and azimuthal phase $e^{im\varphi}$ fixed by the driving QNM.

\section{Shadow Ringing} \label{sec5}
We now illustrate the analytic predictions of the shadow-ringdown framework. Throughout this section, we normalize the screen radius by the static Schwarzschild value $R_{\rm Schw}$ (for a distant observer), and we parameterize the boundary as
\begin{equation}
(\alpha,\beta)=R(\varphi,t),(\cos\varphi,\sin\varphi),
\qquad
\hat R(\varphi,t)\equiv \frac{R(\varphi,t)}{R_{\rm Schw}}.
\end{equation}
In the linear regime derived previously, the shadow boundary admits the mode expansion
\begin{equation}
\hat R(\varphi,t)=1+\varepsilon\sum_{\ell m}\left[\mathcal{T}_{\ell m}\,e^{-i\omega_{\ell m} t+i m\varphi}+\mathcal{T}^{*}_{\ell m}\,e^{+i\omega^{*}_{\ell m} t-i m\varphi}\right]+\mathcal{O}(\varepsilon^{2}),
\end{equation}
where $\omega_{\ell m}=\omega_{\ell m}^{\rm R}-i\omega_{\ell m}^{\rm I}$ and $\mathcal{T}_{\ell m}$ are the photon sphere transfer coefficients obtained in the previous section. When a single mode dominates (say $(\ell,m)$), the perturbation reduces to
\begin{equation} \label{eq_delta_R-hat}
\delta \hat R(\varphi,t)\simeq 2\varepsilon|\mathcal{T}_{\ell m}|e^{-\omega_{\ell m}^{\rm I} t}\,
\cos\left(\omega_{\ell m}^{\rm R} t+m\varphi+\arg\mathcal{T}_{\ell m}\right),
\end{equation}
which makes the geometry, frequency content, and damping entirely explicit.

\begin{figure}
  \centering
  \includegraphics[width=0.6\linewidth]{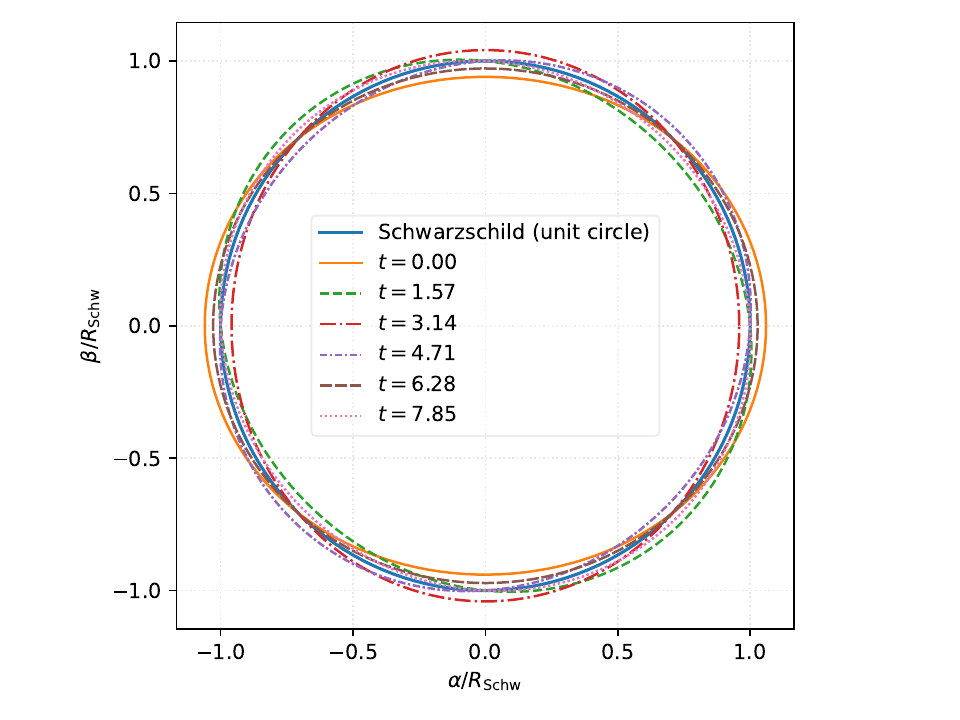}
  \caption{Example instantaneous boundary displacement $\delta R/R_0$ produced by a \emph{single} underlying QNM mode.
Unless stated otherwise, the background is Schwarzschild and we use the leading-order transfer law \eqref{eq_met_pert_R(phi,t)}.
For this visualization we take a representative $(\ell,m)=(2,2)$ perturbation in the polar sector,
$\delta R/R_0 = \epsilon\,\Re\!\left[T^{(\mathrm{pol})}_{22}\,e^{-i\omega t_{\rm obs}} e^{i m\varphi}\right]$,
with time measured in units of the oscillation phase $\tau=\omega_R t_{\rm obs}$ (so the shown phases are $\tau_k=k\pi/2$).
The observer-frame orientation is specified by the Euler angles $(\alpha,\iota,\gamma)$ used in the Wigner-$D$ rotation (set here to $(\alpha,\iota,\gamma)=(\cdots)$).
The plotted amplitude is rescaled by a constant factor (not affecting mode content) for visual clarity.}
  \label{fig1}
\end{figure}

Figure \ref{fig1} displays an oscillatory distortion of the boundary at frequency $\omega_{\rm Re}$ with exponentially shrinking amplitude set by $\omega_{\rm Im}$. The $m$-fold angular symmetry of the deformation is manifest.
As we see, the shadow boundary is displaced by a standing pattern with azimuthal periodicity m and temporal frequency $\omega_{\rm Re}$. The amplitude envelope decays as $e^{-\omega_{\rm Im} t}$, consistent with the ringdown. Because $\hat R$ is normalized by $R_{\rm Schw}$, the leading shape information is disentangled from the absolute scale. For a Schwarzschild background, the selection rules derived earlier imply that only modes allowed by the photon sphere coupling contribute; choosing a single dominant $(\ell,m)$ reproduces the clean $m$-lobed morphology in Eq. \eqref{eq_delta_R-hat}.

\begin{figure}
  \centering
  \includegraphics[width=0.6\linewidth]{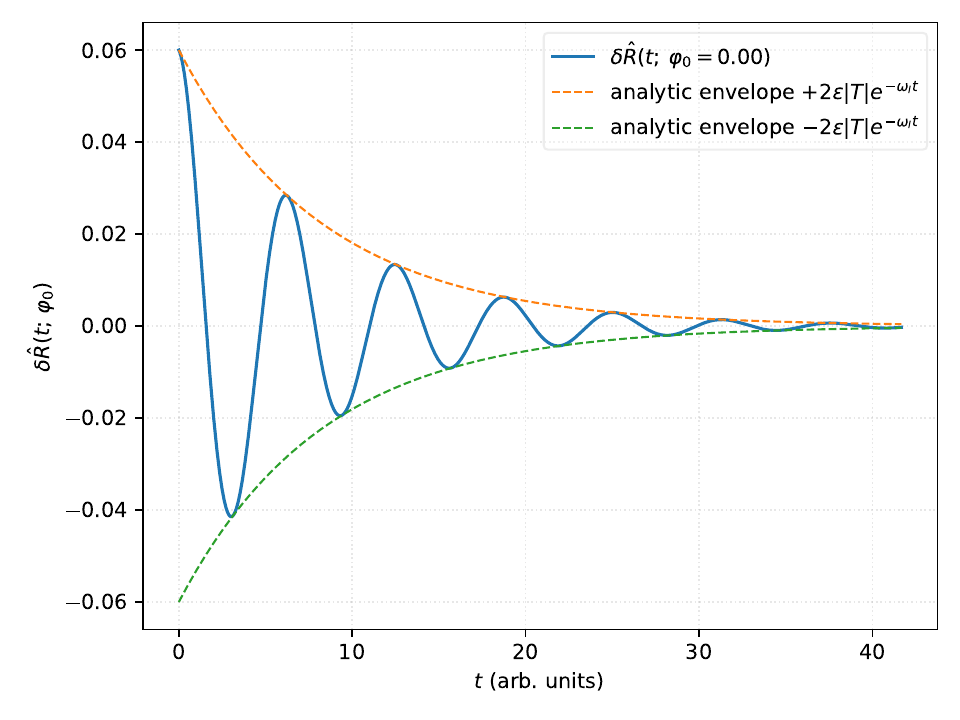}
  \caption{Time evolution of the Fourier coefficients $\hat{R}_m(t_{\rm obs})$ defined in \eqref{eq_R_m(t)}.
The source-frame perturbation is taken to have pure $m_\star=2$ content at fixed $\ell$ (here $\ell=2$), with complex frequency
$\omega=\omega_R+i\omega_I$ and damping factor $e^{\omega_I t_{\rm obs}}$ included.
For an equatorial line of sight ($\iota=0$) the leading-order selection rule implies that only $m=0$ and $m=2$ appear at ${\cal O}(\epsilon)$;
higher harmonics (e.g. $m=4$) enter at ${\cal O}(\epsilon^2)$. We use $\epsilon=\cdots$ and the master-field normalization stated in Sec. \ref{sec4.1}}
  \label{fig2}
\end{figure}

Next, we fix an angle $\varphi=\varphi_{0}$ on the screen and plot the scalar time series
\begin{equation} \label{eq_delta_R-hat2}
\delta \hat R(t,\varphi_0)\equiv \hat R(\varphi_0,t)-1
= \sum_{\ell m} \varepsilon\left(\mathcal{T}_{\ell m} e^{-i\omega_{\ell m} t+i m\varphi_0}+\text{c.c.}\right)+\mathcal{O}(\varepsilon^{2}).
\end{equation}
For a single dominant mode, Eq. \eqref{eq_delta_R-hat2} reduces to a damped cosine with a phase shift $(m\varphi_0+\arg\mathcal{T}_{\ell m})$. Overlay the analytic envelope $( \pm 2\varepsilon |\mathcal{T}_{\ell m}| e^{-\omega_{\ell m}^{\rm I} t})$. See Fig. \ref{fig2}.
Here, at fixed $\varphi_0$, the shadow displacement acts as a single-pixel ringdown seismometer. In the single-mode limit, fitting Eq. \eqref{eq_delta_R-hat2} yields $(\omega_{\rm Re},\omega_{\rm Im})$ directly, while the $\varphi_0$-dependence of the phase isolates $m$. With multiple active modes, Eq. \eqref{eq_delta_R-hat2} is a short sum of damped sinusoids; the relative phases test the selection rules and the predicted phases of $\mathcal{T}_{\ell m}$.

\begin{figure}
  \centering
  \includegraphics[width=0.6\linewidth]{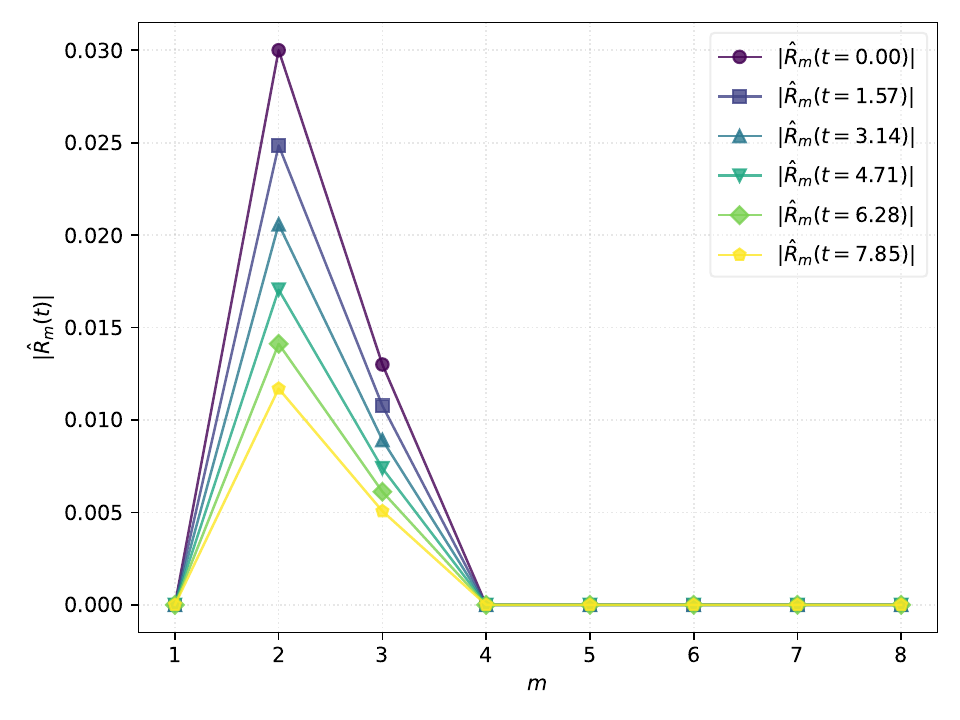}
  \caption{Screen-frame mode mixing for a tilted line of sight $\iota\neq 0$.
The underlying perturbation has pure source-frame $m_\star=2$ at fixed $\ell$ (here $\ell=2$), and the observer frame is related by the
Wigner rotation $D^\ell_{m m_\star}(\alpha,\iota,\gamma)$.
We plot $\hat{R}_m(t_{\rm obs})$ for the specific angles $(\alpha,\iota,\gamma)=(\cdots)$ and parameters $(\epsilon,\omega)=(\cdots)$.
As expected, all $|m|\le \ell$ are populated with weights controlled by $|D^\ell_{m m_\star}|$.}
  \label{fig3}
\end{figure}

To make the selection rule visible, we fix any time $t$, compute the azimuthal Fourier coefficients
\begin{equation} \label{eq_R_m(t)}
    \hat R_m(t)\equiv \frac{1}{2\pi}\int_{0}^{2\pi} d\varphi\, \hat R(\varphi,t)\, e^{-i m\varphi},
\end{equation}
which, to linear order, satisfy
\begin{equation}
    \hat R_m(t)\simeq \sum_{\ell}\varepsilon\mathcal{T}_{\ell m} e^{-i\omega_{\ell m} t} + \mathcal{O}(\varepsilon^{2}).
\end{equation}
Figure \ref{fig3} plots the magnitudes of the azimuthal Fourier coefficients \(|\hat R_m(t)|\) against \(m\) for several times \(t\) (e.g., \(t=0.00,1.57,3.14,4.71,6.28,7.85\)), thereby converting the boundary's geometry into a harmonic fingerprint. Multiple time snapshots are overlaid to emphasize a common exponential decay pattern compatible with a QNM origin, while \(m\)'s forbidden by symmetry appear near zero across all times.
We see that to the linear order, \(\hat R_m(t)\propto \sum_{\ell}\mathcal{T}_{\ell m}e^{-i\omega_{\ell m}t}\), so the set of populated \(m\)'s is time‐independent and directly encodes the selection rules tied to the photon sphere coupling (parity, spin weight, and geometry). The relative heights within that set reflect \(|\mathcal{T}_{\ell m}|\), while a uniform vertical shift with time across the active bars indicates a common damping rate \(\omega_{\rm Im}\) for a dominant family of modes. In cases with multiple contributions at the same \(m\), coherent addition of complex amplitudes produces constructive or destructive interference, which is visible as time-dependent beating in \(|\hat R_m(t)|\) at fixed \(m\).

Figs. \ref{fig1}-\ref{fig3} jointly verify the internal logic of the framework and suggest a practical inference pathway that remains purely analytic at leading order: (i) use Fig. \ref{fig1} to identify the dominant \(m\) (visual morphology) and to sanity-check the smallness of the deformation relative to the unit circle; (ii) extract \((\omega_{\rm Re},\omega_{\rm Im})\) from Fig. \ref{fig2} via a damped-cosine fit; and (iii) confirm the selection rules and estimate relative couplings \(|\mathcal{T}_{\ell m}|\) from Fig. \ref{fig3} by comparing bar heights at a fixed reference time. Because each step relies on the same first-order expansion, agreement across panels is a stringent, over-constrained test. Deviations (e.g., phase drifts that cannot be absorbed into \(\arg\mathcal{T}_{\ell m}\), or \(m\)-content inconsistent with the symmetry analysis) would directly point either to higher-order corrections (\(\mathcal{O}\varepsilon^2)\)) or to additional physical effects (finite-distance aberration, motion, or nonseparable perturbations) discussed earlier in the paper.

These visualizations deliberately isolate the theory's leading, mode-resolved predictions. At higher accuracy, one should account for: (a) subdominant modes with nearby frequencies, which can introduce beating in Fig. \ref{fig2} and time-dependent interference within a given \(m\) in Fig. \ref{fig3}; (b) observer motion or finite-distance effects, which primarily rescale and aberrate the circle while weakly mixing neighboring \(m\)'s; and (c) mild gauge artefacts, which are suppressed on the screen but may alter intermediate representations. None of these caveats obscures the core signatures highlighted here, an \(m\)-fold boundary oscillation at \(\omega_{\rm Re}\) with damping \(\omega_{\rm Im}\) and a harmonic spectrum controlled by the photon‐sphere transfer coefficients, so the visuals form a compact, falsifiable summary of shadow ringing.

\section{Conclusion} \label{sec6}
We have developed a compact, analytic description of shadow ringing, the coherent modulation of a black hole's shadow boundary during ringdown. By treating the shadow as a dynamical separatrix of the null geodesic flow and working to first order in a weakly time-dependent metric, we obtained a local, gauge-insensitive transfer law that maps a single QNM perturbation $h_{\mu\nu}\propto e^{-i\omega t}Y_{\ell m}$ to the instantaneous boundary displacement $\delta R(\varphi,t)$. This law shows that the boundary oscillates at the QNM frequency and damps at the QNM rate, that the azimuthal structure on the screen directly reflects the spherical-harmonic index $m$, and that parity-dependent selection rules sharply constrain which channels contribute at leading order. The entire effect is anchored at the photon sphere, making it both theoretically transparent and observationally clean: a Fourier analysis of the boundary alone isolates the active $m$ and returns the complex frequency, independent of radiative-transfer assumptions. The figures (see Figs. \ref{fig1}-\ref{fig3}) confirm these qualitative and quantitative features and clarify how symmetry governs the visible pattern.

While these perturbations are dynamical and (in principle) observationally accessible, an order-of-magnitude estimate indicates that detecting them with
current horizon-scale VLBI imaging would be challenging.
At leading order one expects an angular modulation of order
$\delta\theta \sim \epsilon\,\kappa\,\theta_{\rm sh}$, where $\theta_{\rm sh}$ is the angular shadow radius and $\kappa\equiv\sup|T^{(s)}_{\ell m}|={\cal O}(1)$.
Using the EHT-inferred ring diameter for Sgr A* (about $51.8\,\mu{\rm as}$) and M87* (tens of $\mu{\rm as}$), a conservative $\epsilon\sim 10^{-2}$ gives
$\delta\theta$ at the level of $\lesssim 10^{-1}\,\mu{\rm as}$ for $\kappa\sim 0.3$, well below present-day $\sim(10$--$20)\,\mu{\rm as}$ imaging resolution.
Moreover, for a representative Schwarzschild $\ell=2$ gravitational QNM one has period $T\simeq (2\pi/\omega_R)\,M \approx 16.8\,M$ and damping time
$\tau\simeq (1/|\omega_I|)\,M\approx 11.2\,M$, corresponding to minutes for Sgr~A* but days for M87*, implying very different cadence requirements.
A realistic detection strategy would therefore likely require (i) substantially improved angular resolution and photon-ring sensitivity (e.g. extended baselines or space-VLBI),
and (ii) analysis methods that isolate the boundary response from accretion-flow variability. We leave a detailed observational forecast to future work.

The present results provide a minimal, falsifiable prediction for horizon-scale imaging during the ringdown epoch: an $m$-resolved, exponentially damped boundary modulation locked to QNM frequencies. Several natural extensions follow from this groundwork. The most pressing is the generalization to Kerr. Recasting the transfer law with Teukolsky variables and metric reconstruction should reveal frame-dragging signatures, prograde/retrograde splitting, and a richer $m$-structure produced by spherical-photon-orbit families; these ingredients will introduce controlled phase drifts and beat patterns in $\delta R(\varphi,t)$ beyond the static-axisymmetric limit. A second direction is multimode and overtone content. Because the present framework is linear, superpositions add straightforwardly and predict beating both in the azimuthal harmonics of the boundary and in their temporal envelopes, opening a boundary-only route to multimode spectroscopy aligned with gravitational-wave analyses. A third direction is to push beyond first order. At second order, one expects mild frequency renormalization, weak mode-mode coupling, and $m$-mixing even within a fixed parity; extending the gauge argument and separatrix calculation accordingly would quantify the range of validity of the linear predictions and identify clean higher-order diagnostics.

Observer systematics can be incorporated without changing the principal conclusions. Large but finite observation distances mainly rescale the boundary, while small boosts induce predictable aberrations; both effects can be folded into the present mapping with subleading corrections that preserve the selection rules. Weak dispersive media provide another interesting axis: a slowly varying plasma index would introduce controlled chromatic shifts of the separatrix, yielding frequency-dependent shadow ringing that can serve as a systematic check in dynamic, multi-band imaging. Finally, it will be valuable to confront these predictions with numerical spacetimes. Applying the boundary extraction to snapshots from numerical-relativity ringdowns can validate the transfer coefficients, illuminate higher-order or nonseparable effects when present, and help design boundary-focused diagnostics for future observations.

In sum, the analysis isolates a photon sphere-controlled imprint in the image plane that is both theoretically crisp and observationally accessible. Pursuing the Kerr extension, quantifying modest nonlinearities, and testing the framework on numerical ringdowns are the next steps toward a unified inference of black-hole properties from joint gravitational-wave and horizon-scale imaging data.

\acknowledgments
R. P. would like to acknowledge networking support of the COST Action CA21106 - COSMIC WISPers in the Dark Universe: Theory, astrophysics and experiments (CosmicWISPers), the COST Action CA22113 - Fundamental challenges in theoretical physics (THEORY-CHALLENGES), the COST Action CA21136 - Addressing observational tensions in cosmology with systematics and fundamental physics (CosmoVerse), the COST Action CA23130 - Bridging high and low energies in search of quantum gravity (BridgeQG), and the COST Action CA23115 - Relativistic Quantum Information (RQI) funded by COST (European Cooperation in Science and Technology). R. P. would also like to acknowledge the funding support of SCOAP3.

\appendix
\section{Osculating formalism, forcing, and instantaneous critical curve} \label{ApdXA}
We collect here the osculating-geodesic formalism used in Sec. \ref{sec3.5} and its relation to the instantaneous critical curve that determines the shadow boundary.

\subsection{Hamiltonian and osculating constants}
We work with the null Hamiltonian
\begin{equation}
H(x,p)=\frac{1}{2}\,g^{\mu\nu}(x)k_\mu k_\nu,\qquad
k_\mu \equiv p_\mu,
\end{equation}
with the metric split into a Schwarzschild background and a small perturbation,
\begin{equation}
g^{\mu\nu}=g^{(0)\mu\nu}-\varepsilon h^{\mu\nu},\qquad
g_{\mu\nu}=g^{(0)}_{\mu\nu}+\varepsilon h_{\mu\nu},
\end{equation}
and all raising/lowering done with $g^{(0)}_{\mu\nu}$, so that
\begin{equation}
h_{\mu\nu}=g^{(0)}_{\mu\alpha}g^{(0)}_{\nu\beta}h^{\alpha\beta}.
\end{equation}
We denote
\begin{equation}
H_0(x,p)\equiv \frac{1}{2}g^{(0)\mu\nu}k_\mu k_\nu,\qquad
\delta H \equiv H-H_0 = -\frac{1}{2}h^{\mu\nu}k_\mu k_\nu ,
\end{equation}
where $\delta H=\mathcal{O}(\varepsilon)$.

Hamilton's equations read
\begin{equation}
\dot x^\mu=\frac{\partial H}{\partial k_\mu}=g^{\mu\nu}k_\nu,\qquad
\dot k_\mu=-\partial_\mu H,
\end{equation}
where $\dot{}\equiv d/d\lambda$ and $\lambda$ is an affine parameter.
Expanding to first order,
\begin{equation}
\dot k_\mu
=-\partial_\mu H_0 - \partial_\mu \delta H
= \dot k^{(0)}_\mu + \frac{1}{2}\partial_\mu h^{\alpha\beta}k_\alpha k_\beta
+ \mathcal{O}(\varepsilon^2),
\end{equation}
with $\dot k^{(0)}_\mu=-\partial_\mu H_0$ the background geodesic equation.

In the osculating-picture, the true null ray is described as a Schwarzschild null geodesic with constants of motion that slowly evolve under the forcing induced by the perturbation. For the present application we only need the energy and azimuthal angular momentum,
\begin{equation}
E(\lambda)\equiv -k_t,\qquad
L_z(\lambda)\equiv k_\phi,
\end{equation}
which are constant in the background but acquire $\mathcal{O}(\varepsilon)$ variations. Define their perturbations
\begin{equation}
\delta E(\lambda)\equiv E(\lambda)-E_0,\qquad
\delta L_z(\lambda)\equiv L_z(\lambda)-L_{z0},
\end{equation}
with $E_0$ and $L_{z0}$ the background constants on the reference geodesic. From the above,
\begin{equation}
\dot{\delta E}\equiv -\dot k_t
=-\frac{1}{2}(\partial_t h^{\alpha\beta})k_\alpha k_\beta,
\qquad
\dot{\delta L}_z\equiv \dot k_\phi
=+\frac{1}{2}(\partial_\phi h^{\alpha\beta})k_\alpha k_\beta .
\end{equation}
These are the osculating forcing laws for $E$ and $L_z$.

We introduce the impact parameter
\begin{equation}
b(\lambda)\equiv \frac{L_z(\lambda)}{E(\lambda)} ,
\end{equation}
which controls whether the photon is captured or escapes. Differentiating,
\begin{equation}
\dot b
=\frac{d}{d\lambda}\left(\frac{L_z}{E}\right)
=\frac{\dot L_z E - L_z\dot E}{E^2}
=\frac{1}{2E^2}
\left(\partial_\phi h^{\alpha\beta}-b\,\partial_t h^{\alpha\beta}\right)k_\alpha k_\beta .
\end{equation}
This is the master osculating law for the impact parameter used in the main text, expressed entirely in terms of the raised perturbation $h^{\mu\nu}$ and the covariant momentum $k_\mu$ of the underlying Schwarzschild null generator.

\subsection{Instantaneous critical curve from the perturbed Hamiltonian}
At fixed $(t,\theta,\phi)$, the instantaneous photon separatrix (shadow boundary) is defined on the perturbed spacetime as the locus of circular null orbits that separate capture from escape. In Hamiltonian form, for equatorial motion, it is determined by
\begin{equation}
H(r,p_r=0,b, t,\theta,\phi)=0,\qquad
\partial_r H(r,p_r=0,b, t,\theta,\phi)=0.
\end{equation}
We expand these conditions about the unperturbed circular photon orbit of Schwarzschild, located at
\begin{equation}
r_c^{(0)}=3M,\qquad
b_c^{(0)}=\frac{L_z}{E}\Big|_{c}=3\sqrt{3}\,M.
\end{equation}
We restrict to the equatorial plane, where the separatrix is generated for Schwarzschild.

Write the full circular solution as
\begin{equation}
r_c=r_c^{(0)}+\delta r_c,\qquad
b_c=b_c^{(0)}+\delta b_c,
\end{equation}
with $\delta r_c,\delta b_c=\mathcal{O}(\varepsilon)$.
Set $p_r=0$ and view the Hamiltonian as a function of $(r,b)$ at fixed $E$,
\begin{equation}
H_0(r,p_r=0,b)
=\frac{1}{2}\left(g^{tt}E^2+g^{\phi\phi}L_z^2\right)
=\frac{E^2}{2}\left(-\frac{1}{f}+\frac{b^2}{r^2}\right),
\qquad
f(r)=1-\frac{2M}{r}.
\end{equation}
The circular null orbit of the background satisfies
\begin{equation}
H_0(r_c^{(0)},0, b_c^{(0)})=0,\qquad
\partial_r H_0(r_c^{(0)},0, b_c^{(0)})=0,
\end{equation}
which gives $r_c^{(0)}=3M$ and $b_c^{(0)}=3\sqrt{3}\,M$.

Now linearize the full conditions
\begin{equation}
H(r_c+\delta r_c,0, b_c+\delta b_c)=0,\qquad
\partial_r H(r_c+\delta r_c,0, b_c+\delta b_c)=0
\end{equation}
in $\delta r_c$, $\delta b_c$, and $\delta H$. Terms such as $\delta r_c\,\partial_r\delta H$ are second order and can be dropped. Using $H=H_0+\delta H$ and the background circular-orbit relations, we obtain
\begin{align}
0&=\delta H + (\partial_r H_0)_c\,\delta r_c + (\partial_b H_0)_c\,\delta b_c,
\\
0&=\partial_r\delta H + (\partial_{rr}H_0)_c\,\delta r_c + (\partial_{rb}H_0)_c\,\delta b_c,
\end{align}
where a subscript $c$ denotes evaluation at the background circular orbit $(r_c^{(0)},b_c^{(0)})$ with $p_r=0$.
Because $(\partial_r H_0)_c=0$ but $(\partial_b H_0)_c\neq 0$, the $2\times2$ system is non-degenerate and both equations must be kept.

The relevant Schwarzschild derivatives at $r_c^{(0)}=3M$, $b_c^{(0)}=3\sqrt{3}\,M$ are
\begin{equation}
(\partial_b H_0)_c = \frac{\sqrt{3}\,E^2}{3M},\qquad
(\partial_{rb} H_0)_c = -\frac{2\sqrt{3}\,E^2}{9M^2},\qquad
(\partial_{rr} H_0)_c = -\frac{E^2}{M^2}.
\end{equation}
The first linearized equation then gives directly
\begin{equation}
(\partial_b H_0)_c\,\delta b_c + \delta H_c = 0,
\end{equation}
so that
\begin{equation}
\delta b_c(t,\varphi)
= -\frac{\delta H_c}{(\partial_b H_0)_c}
= -\frac{3M}{\sqrt{3}E^2}\,\delta H_c(t,\varphi),
\end{equation}
and hence
\begin{equation}
\frac{\delta b_c(t,\varphi)}{b_c^{(0)}}
=-\,\frac{\delta H_c(t,\varphi)}{3E^2}.
\end{equation}
Here
\begin{equation}
\delta H_c(t,\varphi)
=-\frac{1}{2}h^{\mu\nu}(t,r,\theta,\phi)k_\mu k_\nu\Big|_{r=3M,\ \theta=\frac{\pi}{2},\ \phi=\varphi},
\end{equation}
with $k_\mu=(-E,0,0,L_z)$ the covariant momentum of the circular background photon orbit, so that only the $tt$, $t\phi$, and $\phi\phi$ components of $h^{\mu\nu}$ enter. The second linearized equation can be used to solve for $\delta r_c$ in terms of $\delta H_c$ and $\partial_r\delta H_c$, but the shadow radius depends only on $\delta b_c$.

For a distant, static observer, the screen radius of a null ray asymptotically approaches the impact parameter, so that $R=b$ at leading order in $M/r_{\rm obs}$. Taking $R_0=b_c^{(0)}$ for the unperturbed critical curve, the fractional shift of the instantaneous critical curve on the screen is therefore
\begin{equation}
\frac{\delta R(\varphi,t_{\rm obs})}{R_0}
=\frac{\delta b_c(\varphi,t_{\rm obs})}{b_c^{(0)}}
=-\,\frac{\delta H_c(\varphi,t_{\rm obs})}{3E^2}
=\frac{1}{6E^2}h^{\mu\nu}(t_{\rm obs},r,\theta,\phi)\,k_\mu k_\nu\Big|_{r=3M,\ \theta=\frac{\pi}{2},\ \phi=\varphi}.
\end{equation}
Because the null dynamics is homogeneous in $E$, we are free to fix the affine normalization so that $E=1$; with this choice,
\begin{equation}
\frac{\delta R(\varphi,t_{\rm obs})}{R_0}
=\frac{1}{6}\,
h^{\mu\nu}(t_{\rm obs},r,\theta,\phi)\,k_\mu k_\nu\Big|_{r=3M,\ \theta=\frac{\pi}{2},\ \phi=\varphi},
\end{equation}
which coincides with the transfer law Eq. \eqref{eq_met_pert_R(phi,t)} quoted in the main text (with the understanding that the physical perturbation is the real part plus its complex conjugate for a QNM mode).

\subsection{Gauge behaviour}
A first-order gauge transformation of the metric perturbation
\begin{equation}
h_{\mu\nu}\to h_{\mu\nu} + 2\nabla_{(\mu}\xi_{\nu)}
\end{equation}
induces
\begin{equation}
h^{\mu\nu}\to h^{\mu\nu} - \nabla^{(\mu}\xi^{\nu)},
\end{equation}
with indices raised using the background Schwarzschild metric. The corresponding change in the Hamiltonian perturbation is
\begin{equation}
\delta(\delta H)
= -\frac{1}{2}(\nabla^{(\mu}\xi^{\nu)})k_\mu k_\nu
= -\frac{1}{2}k^\mu\nabla_\mu(\xi\cdot k),
\end{equation}
where $k^\mu=g^{(0)\mu\sigma}k_\sigma$ and $\xi\cdot k\equiv \xi_\mu k^\mu$.
Thus the change in the transfer law is
\begin{equation}
\delta\!\left(\frac{\delta R}{R_0}\right)
=-\frac{1}{3E^2}\,\delta(\delta H_c)
=\frac{1}{6E^2}\,k^\mu\nabla_\mu(\xi\cdot k)\Big|_{c}.
\end{equation}
For gauge vectors $\xi^\mu$ that are regular on the horizon and decay sufficiently fast at spatial infinity, the quantity $\xi\cdot k$ vanishes at both ends of the photon generator. Since the right-hand side is a total derivative along the background null orbit, these boundary conditions imply that the contribution of the gauge term integrates to zero along the ray, and the local value at the photon sphere is unaffected. In particular, the leading deformation of the critical curve and the mode-selection rules discussed in Secs. \ref{sec3.1}-\ref{sec3.3} are insensitive to such admissible gauge transformations.

\bibliography{ref}

\end{document}